\documentclass[
 reprint, 
amsmath,amssymb,
aps,
physrev,
floatfix
]{revtex4-2}

\usepackage{graphicx}
\usepackage{dcolumn}
\usepackage{float}
\usepackage{bm}
\usepackage[normalem]{ulem} 
\usepackage{subcaption}

\usepackage{color}
\newcommand{\numu}{$\nu_{\mu}$}
\newcommand{\nue}{$\nu_{e}$}

\newcommand{\thetamu}{$\theta_{23}$}
\newcommand{\thetae}{$\theta_{13}$}
\newcommand{\thetasol}{$\theta_{12}$}

\newcommand{\sinsqthetareac}{$\sin^2\theta_{13}$}
\newcommand{\sinsqthetamu}{$\sin^2\theta_{23}$}

\newcommand{\dcp}{$\delta_{\text{CP}}$}

\newcommand\brabar{\raisebox{-4.0pt}{\scalebox{.2}{
\textbf{(}}}\raisebox{-4.0pt}{{\_}}\raisebox{-4.0pt}{\scalebox{.2}{\textbf{)
}}}}
\begin{document}

\preprint{APS/123-QED}

\title{On Precision of the Leptonic Mixing Angle $\theta_{23}$ and \\its Implications for the Flavor Models}
\author{P. T. Quyen$^{1,2,*}$}
\author{S. Cao$^{1}$}%
\thanks{These authors contributed equally to this work. \\Corresponding author cvson@ifirse.icise.vn}
\author{N. T. Hong Van$^{3}$}
\author{Ankur Nath$^{4}$}
\author{T. V. Ngoc$^{5}$}


\affiliation{\vspace*{2mm}
$^1$\textit{Institute for Interdisciplinary Research in Science and Education},\\ 
\it{ICISE, Quy Nhon, Vietnam.\vspace*{2mm}\\}
$^2$\textit{Graduate University of Science and Technology, Vietnam Academy of Science and Technology, Hanoi, Vietnam.}\vspace*{2mm}\\
$^3$\textit{Institute of Physics, Vietnam Academy of Science and Technology, Hanoi, Vietnam}.\\
$^4$\textit{Department of Physics, Namrup College, Assam, India}\\
$^5$\textit{Department of Physics, Kyoto University, Kyoto, Japan}\\
}

\date{\today}

\begin{abstract}
Among three leptonic mixing angles, $\theta_{23}$ angle, which characterizes the fractional contribution of two flavor eigenstates $\nu_{\mu}$ and $\nu_{\tau}$ to the third mass eigenstate $\nu_3$, is known to be the largest but the least precisely measured. The work investigates possible reach of $\theta_{23}$ precision with two forthcoming gigantic accelerator-based long-baseline neutrino experiments, namely Hyper-Kamiokande (T2HK) and DUNE experiments as well as a possible joint analyses of future neutrino facilities. Our simulation yields that each experiment will definitely establish the octant of $\theta_{23}$ angle for all values within 1$\sigma$ parameter interval, while considering the current limitation. However, if the actual value is $0.48\leq \sin^2\theta_{23}\leq 0.54$, it becomes challenging for these two experiments to reject the maximal ($\theta_{23}=\pi/4$) hypothesis and conclude its octant. This octant-blind region can be further explored with the proposed facilities ESSnuSB and a neutrino factory. Accurate determination of the mixing angle $\theta_{23}$, as well as the accuracy of $\delta_{CP}$, is crucial for examining a certain category of discrete non-Abelian leptonic flavor models. Specifically if CP is conserved in leptonic sector, the combined analysis of T2HK and DUNE will rule out the majority of these models. However, if the CP is maximally violated, higher precision of $\delta_{CP}$ is necessary for testing these flavor models.
\end{abstract}
\maketitle

\section{\label{sec:intro}Current understandings of neutrino oscillation parameters}
Observation of neutrino oscillation phenomenon~\cite{fukuda1998evidence, Ahmad:2001an,ahmad2002measurement} revolutionizes particle physics at the dawn of the twenty-first century since its implications of massive neutrinos and leptonic mixing are not adequately explained by the Standard Model of elementary particles. The up-to-date data~\cite{ParticleDataGroup:2024cfk}, with few anomaly exceptions, can be well-described by a three-flavor neutrino model based on a $3\times3$ unitary mixing matrix known as Pontecorvo–Maki–Nakagawa–Sakata (PMNS) matrix~\cite{pontecorvo1968neutrino,maki1962remarks}. The matrix represents the magnitude of the coupling between three neutrino mass eigenstates $(\nu_{1}, \nu_{2}, \nu_{3})$ and three charged-lepton states $(e,\ \mu,\ \tau)$. The PMNS matrix is conventionally parameterized by three mixing angles ($\theta_{12}, \theta_{13}, \theta_{23}$), one Dirac CP-violation phase \dcp, and additional two Majorana CP-violation phases if the neutrinos are Majorana particles. Measurements using neutrino oscillation phenomena, which are not affected by the Majorana phases, enable us to determine the four PMNS oscillation parameters ($\theta_{12}, \theta_{13}, \theta_{23}$, \dcp) and the neutrino mass-squared splittings, represented as $\Delta m^2_{ij} = m^2_i-m^2_j$ where $(i,j)={1,2,3}$. The neutrino oscillation measurements typically involve two types of data samples: (i) the survival or \emph{disappearance} of an $\alpha$-flavor from the neutrino production source, and (ii) the \emph{appearance} of a $\beta$-flavor from the  $\alpha$-flavor neutrino production source. The aforementioned process is observed in the survivals of $\nu_e$ from sun, $\overline{\nu}_{e}$ from the reactors and $\nu_{\mu} (\overline{\nu}_{\mu})$ from the atmospherics and from accelerator-based sources. The \emph{appearances} of $\nu_{e} (\overline{\nu}_{e})$ and $\nu_{\tau} (\overline{\nu}_{\tau})$ from atmospheric and accelerator-based sources of $\nu_{\mu} (\overline{\nu}_{\mu})$ exemplify the later process. The probability of ${\alpha}$-flavor neutrinos with energy $E$ transitioning into ${\beta}$-flavor neutrinos observed at a distance of $L$ in vacuum can be expressed as
 \begin{align*}
P(\nu_{\alpha} \rightarrow \nu_{\beta})=\delta_{\alpha \beta}&-4\sum_{i>j}\Re(U^*_{\alpha i}U_{\beta i} U_{\alpha j} U^*_{\beta j})\sin^2 \Phi_{ij} \nonumber \\
&\pm 2\sum_{i>j} \Im(U^*_{\alpha i}U_{\beta i} U_{\alpha j} U^*_{\beta j})\sin 2 \Phi_{ij}
\end{align*}
where $\Phi_{ij} = \Delta m^2_{ij}\frac{L}{4E}\equiv 1.27\times \Delta m^2_{ij} [\text{eV}^2]\frac{L[\text{km}]}{E[\text{GeV}]}$ and $\pm$ sign is taken for neutrinos and anti-neutrinos, respectively. There are two well-established scales for neutrino mass-squared splittings: $\Delta m^{2}_{21}\sim 7.4\times 10^{-5}\ \text{eV}^{2}/c^{4}$ and $|\Delta m^{2}_{31}|\sim 2.5\times 10^{-3} \ \text{eV}^{2}/c^{4}$. A crucial point to emphasize is that the positive or negative nature of $\Delta m^{2}_{31}$ remains unknown at yet. Therefore, it is currently unknown whether neutrino mass spectrum adheres to the \emph{normal} ordering ($m_3>m_2>m_1$) or \emph{inverted} ordering ($m_2>m_1>m_3$). Experiments T2K~\cite{T2K:2023smv}, NO$\nu$A~\cite{acero2022improved} and Super-Kamiokande (Super-K)~\cite{Super-Kamiokande:2023ahc} individually shows some mild preference to the \emph{normal} mass ordering over the \emph{inverted} one. Combining higher statistical samples from T2K and NO$\nu$A with the reactor-based medium-baseline experiment JUNO~\cite{djurcic2015juno} will be crucial for elucidating this unknown~\cite{Cao:2020ans,Cabrera:2020ksc}. One extra mystery in the PMNS 3-flavor picture is the parameterization-independent amplitude of CP violation, known as leptonic Jarlskog invariance, which is directly related to the CP-violation phase \dcp\ as
\begin{align}
J_{\text{CP}}^{\text{Lepton}} &= \Im[U_{\alpha i}U^*_{\alpha j} U^*_{\beta i} U_{\beta j}]\\ \nonumber
&= \frac{1}{8} \sin2\theta_{12}\sin2\theta_{23}\sin2\theta_{13}\cos\theta_{13}\sin\delta_{\text{CP}}.
\end{align}
 The T2K  experiment~\cite{T2K:2019bcf,T2K:2023smv} has recently presented a significant hint on the non-zero CP-violation phase. However, NO$\nu$A experiment~\cite{acero2022improved} does not exhibit any similar inclination in the available data.  The potential detection of the CP violation before the commencement of the next-generation accelerator-based neutrino experiments Hyper-Kamiokande (T2HK)~\cite{protocollaboration2018hyperkamiokande} and DUNE~\cite{DUNE:2020lwj} make the future joint analysis of T2K and NO$\nu$A~\cite{Cao:2020ans} a fascinating prospect. The final uncertainty pertains to the proximity of the mixing angle \thetamu\ to $\pi/4.$ 
  \begin{table}
    \centering
    \begin{tabular}{l|c|c}
    \hline\hline
    Parameter & Best fit & $3\sigma$ C.L. range \\\hline
    $\sin^{2}\theta_{12}$ & $0.303$ & [0.270, 0.341]\\ 
    $\sin^{2}\theta_{13}(\times 10^{-2})$ & 2.203 & [2.0, 2.4]\\
    $\sin^{2}\theta_{23}$ & 0.572 & [0.406, 0.620]  \\ 
    $\delta_{CP}(^{\circ})$ & 197 & [108, 404] \\
  $\Delta m^{2}_{21} (10^{-5}\text{eV}^{2}/c^4)$ & 7.41 & [6.82, 8.03] \\
   $\Delta m^{2}_{31} (10^{-3}\text{eV}^{2}/c^{4})$ & 2.511 & [2.428, 2.597] \\\hline
    \end{tabular}
    \caption{Global constraints of neutrino oscillation parameters with \emph{normal} mass ordering assumed, taken from  Ref.~\cite{Esteban:2020cvm} with NuFit 5.2 based on data available in November 2022.}
   \label{tab:nuoscpara}
\end{table}
As shown in Table~\ref{tab:nuoscpara}, the current 3$\sigma$ C.L. range of $\sin^2\theta_{23}$ encompasses around 21\% of all possible values. The data strongly support the maximal mixing $\theta_{23}=\pi/4$ hypothesis. The near proximity of the mixing angle \thetamu\ to maximal is an indication of a hidden symmetry between the second and third lepton generations, which are two distinct copies of the irreducible representations of $\text{SU}(2)_{L}$ group. The precise value of $\theta_{23}$ would be an important input to the flavor and neutrino mass models, as illustrated in Ref.~\cite{Mohapatra:2006gs,Gehrlein:2022nss} and references therein. 

This work aims to provide a concise overview of the uncertainty associated with measuring the $\theta_{23}$ mixing angle, estimate the possible reach of $\theta_{23}$, and analyze its implications for a certain category of flavor models.
\section{\label{sec:ambiguity}Ambiguity in measuring the $\theta_{23}$ mixing angle}

The current knowledge of $\theta_{23}$ is mostly derived from neutrino oscillation measurements conducted using two sources: (i) atmospheric neutrinos and (ii) accelerator-based neutrinos. Joint analysis of accelerator-based and atmospheric neutrino sources in MINOS and MINOS+ experiments~\cite{MINOS:2020llm} results in $\sin^{2}\theta_{23} = 0.43^{+0.20}_{-0.04}$. Measurement using the atmospheric neutrino data from Super-K~\cite{Super-Kamiokande:2010orq} yields an interval of $0.41 \leq \sin^{2}\theta_{23} \leq 0.58$ at the 90\% C.L. Both current data from accelerator-based long-baseline T2K~\cite{T2K:2023smv} and  NO$\nu$A~\cite{acero2022improved} favor slightly the upper octant of the mixing angle \thetamu. The current global fit data~\cite{Esteban:2020cvm} using NuFit 5.2 supports the maximal mixing hypothesis at 90\% C.L. The measurements from leading experiments and the global analysis of neutrino oscillation measurements are summarized in Table~\ref{tab:datath23}.  
\begin{table*}
     \centering
\begin{ruledtabular}
 \begin{tabular}{c|c|c|c|c|c|c}
  & T2K & NO$\nu$A & MINOS & Super-K & IceCube & NuFIT 5.2 \\
  \hline
  Best fit $\sin^{2}\theta_{23}$ &$0.561^{+0.019}_{-0.038}$ & $0.57^{+0.03}_{-0.04}$& $0.43^{+0.20}_{-0.04}$& $0.425^{+0.051}_{-0.034}$ &  $0.51\pm 0.05$&$0.572^{+0.018}_{-0.023}$\\
  \hline
  Maximal rej.[$\sigma$] & 1.22 & 1.29 & 0.90 & 1.25 & 0.28 & 1.69 \\
  \hline
  Wrong-octant rej.[$\sigma$] & 1.22 & 0.37 & 0.53& 0.85 & 0 & 0.89 \\
  \hline
   Constrained $\sin^{2}\theta_{13}/10^{-2}$ & $2.18\pm0.07$ & $ 2.10\pm 0.11$ & $2.10\pm0.11$ & $2.10\pm0.11$ & $2.224\pm0.11$ & $2.203\pm0.0575$ \\ 
   \end{tabular}
  \end{ruledtabular}
  \caption{\label{tab:datath23} Current constraints on $\sin^{2}\theta_{23}$ from T2K~\cite{T2K:2023smv}, NO$\nu$A~\cite{NOvA:2021nfi}, MINOS(+)~\cite{MINOS:2020llm}, Super-K~\cite{Super-Kamiokande:2019gzr}, IceCube ~\cite{IceCubeCollaboration:2023wtb} and NUFIT 5.2~\cite{Esteban:2020cvm} with a \emph{normal} mass ordering assumed.
}
\end{table*}
\noindent In long-baseline experiments using accelerator-based neutrino sources like ongoing T2K ~\cite{Abe:2011ks}, and NO$\nu$A~\cite{ayres2007nova}, and forthcoming T2HK ~\cite{protocollaboration2018hyperkamiokande} and DUNE~\cite{DUNE:2020lwj}, the precise $\theta_{23}$ value can be extracted from measurements of \numu ($\bar{\nu_{\mu}}$)-survival probabilities ($P(\nu_{\mu} \xrightarrow{} \nu_{\mu}), P(\overline{\nu}_{\mu} \xrightarrow{} \overline{\nu}_{\mu})$ or called as \emph{disappearance} samples) and/or \emph{appearances} of electron neutrinos from muon neutrinos
($P(\nu_{\mu} \xrightarrow{} \nu_{e}), P(\overline{\nu}_{\mu} \xrightarrow{} \overline{\nu}_{e})$ and known as \emph{appearance} samples).
Eq.~(\ref{eq:probnumu2numu}) describes the survival probability of muon neutrinos around the oscillation maximum $\Phi_{31}\equiv 1.27\times \Delta m^2_{31} [\text{eV}^2]\frac{L[\text{km}]}{E[\text{GeV}]} \approx \pi/2$, which is applicable to T2K, NO$\nu$A, and T2HK experimental setups.
\begin{align} \label{eq:probnumu2numu}
    &P_{\nu_{\mu}\rightarrow \nu_{\mu}} \left( \Phi_{31}\approx \pi/2\right) \nonumber \\
    &= 1 - \left( \cos^4 \theta_{13} \sin^2 2\theta_{23}+\sin^22\theta_{13}\sin^2\theta_{23}\right)\sin^2\Phi_{31} \nonumber \\
&+ \epsilon_m \Phi_{31} \sin 2 \Phi_{31} \left( \cos^2\theta_{12}\sin^22\theta_{23}-\sin^2\theta_{23}J_{123}\cos\delta_{\text{CP}} \right), 
\end{align}
where $\epsilon_m = \frac{\Delta m^2_{21}}{\Delta m^2_{31}}$ and $J_{123}=\sin 2\theta_{12}\sin 2\theta_{23}\sin 2\theta_{13}$. The leading order term of Eq.~(\ref{eq:probnumu2numu}) reveals that the survival probability reaches its minimum at approximately $\sin^2\theta_{23}\approx 0.5\cos^{-2}\theta_{13}= 0.51$, given $\sin^2\theta_{13} = 0.02203$. The survival oscillation probability $P_{\nu_{\mu}\rightarrow \nu_{\mu}}$ exhibits symmetry at this point and results in a discrete octant degeneracy: two discrete values of $\sin^2\theta_{23}$ correspond to identical probabilities. To overcome the octant degeneracy in the \emph{disappearance} channels, one approach is to utilize the sample of \emph{appearance} data. Eq.~(\ref{eq:probnumu2nue}) provides the approximate \nue\  \emph{appearance} probability around the oscillation maximum $\Phi_{31}\approx \pi/2$ in vacuum.
\begin{align} \label{eq:probnumu2nue}
    P_{\nu_{\mu}\rightarrow \nu_{e}}& \left( \Phi_{31}\approx \pi/2\right) \nonumber = \sin^2\theta_{23}\sin^22\theta_{13}\sin^2\Phi_{31} \nonumber \\
&+\epsilon_m\Phi_{31}\sin\Phi_{31}J_{123}\cos (\Phi_{31}+\delta_{CP}) ,
\end{align}
The leading term of the $\nu_{e}$ \emph{appearance} probability is dertermined by the factor \sinsqthetamu, making it highly responsive to the octant of \thetamu. In addition, Eq.~(\ref{eq:probnumu2nue}) shows that the magnitude of this flavor transition is modulated by both the poorly-established \dcp\ and unknown neutrino mass ordering. Nevertheless, when both $\nu_{e}$ \emph{appearance} and $\overline{\nu}_{e}$ \emph{appearance} are measured with rather high statistical significance, the octant resolving has a marginal influence of the actual value of \dcp. Besides, it is found that the relative difference in the probabilities of the $\nu_{e}$ ($\overline{\nu}_{e}$) \emph{appearance} between the \emph{normal} and \emph{inverted} mass ordering is largely unaffected by the actual value of \sinsqthetamu\ but only the true value of \dcp. Consequently, the precise measurement of \thetamu\ does not depend significantly on the neutrino mass ordering.

\begin{figure}
\includegraphics[width=0.48\textwidth]{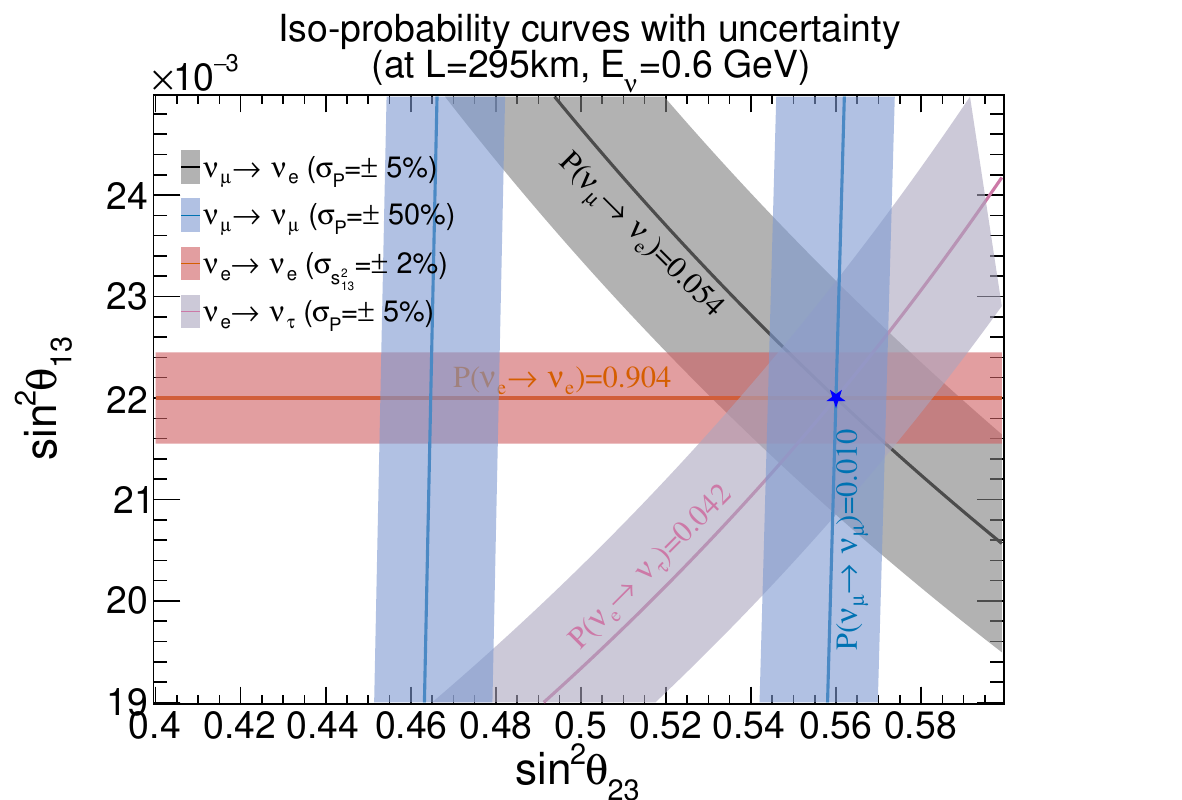}
\caption{\label{fig:th23vsth13} Schematic view of addressing the ambiguity of \thetamu\ is presented with the iso-probability curves in the (\sinsqthetamu,\ \sinsqthetareac) parameter space. The error band for each curve is obtained with a predefined uncertainty of probability $\sigma_{\text{p}}$. This includes $\nu_{\mu}\rightarrow\nu_{\mu}$, $\nu_{\mu}\rightarrow\nu_{e}$, $\nu_{\mu}\rightarrow\nu_{\tau}$, and $\nu_{e}\rightarrow\nu_{e}$ transitions. The true values of parameters are set to \sinsqthetamu= 0.56 and \sinsqthetareac= 0.02203 from NuFIT 5.2.} 
\end{figure}

A relevant parameter which significantly influences the determination of the \thetamu\ octant in case of non-maximal mixing is the precision value of \thetae. While the observation of $\nu_{\mu}\rightarrow \nu_e$ transition by T2K~\cite{T2K:2013ppw} established the non-zero value of \thetae, the most accurate measurement of this mixing angle is obtained from the reactor-based experiments~\cite{ParticleDataGroup:2024cfk}. From Eq.~(\ref{eq:probnumu2nue}), the analytical formula to describe the continuous degeneracy between \thetamu\ and \thetae\ can be formulated as 
\begin{align}
\sin2\theta^{\text{true}}_{13}\sin\theta^{\text{true}}_{23}=\sin2\theta^{\text{cloned}}_{13}\sin\theta^{\text{cloned}}_{23}
\end{align}
where $\left(\theta^{\text{true}}_{13},\theta^{\text{true}}_{23}\right)$ are true values and $\left(\theta^{\text{cloned}}_{13},\theta^{\text{cloned}}_{23}\right)$ are cloned solutions for a given $\nu_{\mu}\rightarrow \nu_e$ probability. This degeneracy suggests that the precision of the later will have a significant impact on the precision, as well as the octant resolving, of the former. Indeed, the effect has been observed in T2K data~\cite{T2K:2023smv} where the data with and without reactor constraint on \thetae\ exhibits different preference for octant of \thetamu. The reason is that T2K data itself favor a greater value of \thetae\ than the stringent constraint obtained from the reactor-based measurements. We provide a summary of the possible measurements in Fig.~\ref{fig:th23vsth13}, which are displayed on the parameter space of \thetamu-\thetae, in order to clarify the uncertainty in determining the mixing angle \thetamu. For illustrative purpose, the true value of \sinsqthetamu\ is set at 0.56 and \sinsqthetareac\ is set at 0.02203. Using measurement with the $\nu_{\mu}\rightarrow \nu_{\mu}$ data channel, two distinct solutions will be obtained, one of which will be cloned in the lower octant. The curves of $\nu_{\mu}\rightarrow \nu_{e}$ and  $\nu_{e}\rightarrow \nu_{\tau}$ iso-probabilities are orthogonal since the leading terms of the probabilities are proportional to $\sin^2\theta_{23}\sin^22\theta_{13}$ for the former and  $\cos^2\theta_{23}\sin^22\theta_{13}$ for the later. Measurements made using $\nu_e\rightarrow\nu_e$ or $\overline{\nu}_e\rightarrow \overline{\nu}_e$ are marginally sensitive to \thetamu\ but they are crucial for accurately measuring the \thetae\ value and hence valuable for resolving the octant of \thetamu\ in case of non-maximal mixing. Recent measurements~\cite{Esteban:2020cvm} have yielded a precision of 2.6\% on $\sin^{2}\theta_{13}$, corresponding to 1.3\% uncertainty of $\theta_{13}$. Furthermore, the value of $\sin^{2}\theta_{13}$ can be improved to $1\%$ (or 0.5\% uncertainty of $\theta_{13})$~\cite{Zhang:2022zoc}. The $\theta_{23}$ sensitivity will be calculated using these two constraints. 

It is important to point out that the primary purpose of the \thetamu\ measurements is to quantitatively assess the promximity of \thetamu\ to the maximal mixing. This, as discussed in Sec.~\ref{sec:flavormodel}, has considerable implications for testing the flavor models. From an empirical standpoint, this entails two consecutive hypothesis tests: (i) excluding the maximal mixing, and (ii) identifying the octant if non-maximal mixing is detected. The former can be quantitatively evaluated using statistics at a given $\theta_{23}^{t}$ in the vector of relevant parameter $\vec{o}$:
\begin{equation*} \label{eq:chiExcludeMax}
    \Delta \chi^{2}_1(\vec{o};\theta_{23}^{t}) = \chi^{2}_{\text{min.}}(\vec{e}; \theta_{23}=\pi/4) - \chi^{2}_{\text{min.}} (\vec{e};\theta_{23}^{\text{b.f.}})
\end{equation*}
where ($\vec{e}; \theta_{23}^{\text{b.f.}}$) is the best-fit parameter vector, which ideally equals to  $\vec{o}$ and $\theta_{23}^{\text{b.f.}}=\theta_{23}^{t}$. The statistical significance for the latter test is defined as the relative difference of $\chi^2$ between the wrong-octant (W.O.) and the true-octant (T.O.) in the \thetamu\ parameter space: 
\begin{equation*} \label{chioctant}
    \Delta \chi^{2}_2(\vec{o};\theta_{23}^{t}) = \chi^{2}_{\text{min.}}(\vec{e}; \theta_{23}^{\text{b.f.}} \text{in W.O.}) - \chi^{2}_{\text{min.}} (\vec{e};\theta_{23}^{\text{b.f.}} \text{in T.O.})
\end{equation*}
Due to the continuity of the neutrino oscillation probability as function of \thetamu, the statistical significance to exclude the maximal mixing is greater than that of excluding the wrong-octant. In the long-baseline accelerator-based experiments, the $\overset{\brabar}{\nu}_{\mu}\rightarrow \overset{\brabar}{\nu}_{\mu}$ \emph{disappearance} samples are the more significant data sample for addressing the first hypothesis test, unless the actual \sinsqthetamu\ exhibits a small deviation from the maximal mixing towards the higher octant. In contrast, the $\overset{\brabar}{\nu}_{\mu}\rightarrow \overset{\brabar}{\nu}_{e}$ \emph{appearance} samples are crucial for resolving octants in case of non-maximal mixing. 

\section{\label{sec:exp} Experimental specifications of T2HK, DUNE, ESSnuSB and Neutrino Factory}
In this section, we detail the experimental specifications for simulating the forthcoming accelerator-based long-baseline neutrino experiments T2HK and DUNE, along with proposed facilities ESSnuSB and Neutrino Factory.
\subsection*{T2HK and DUNE}
Hyper-Kamiokande~\cite{protocollaboration2018hyperkamiokande}, a next-generation water Cherenkov detector that is approximately 8.4 times larger than its predecessor Super-Kamiokande, is a massive observatory under construction aimed at elucidating the mysteries of neutrinos, searching for proton decay, and investigating other new physics phenomena. This observatory is capable of detecting neutrinos from diverse natural (solar, atmospheric, astrophysical) and artificial (reactor, accelerator) sources. The Tokai to Hyper-Kamiokande (T2HK) program employs an intense and highly pure source of muon neutrinos and anti-neutrinos generated by the J-PARC accelerator, located 295 km away and at an angle of $2.5^{\circ}$ from the average beam axis. This experimental setup is to achieve a narrow neutrino energy spectrum and enhance sensitivity to the relevant parameters, particularly CP-violation phase. The experiment is scheduled to commence data collection in 2027 and will continue for a duration of 10 years, equivalent to an exposure of $2.2\times 10^{22}$ protons-on-target (POT). This enables T2HK to collect approximately 10,000 $\nu_{\mu} (\overline{\nu}_{\mu})$ events and 2,000 $\nu_{e} (\overline{\nu}_{e})$ events, hence allowing T2HK to exclude CP conserving values ($\delta_{CP} = 0, \pm\pi)$ at $5\sigma$ or greater C.L. for around 60\% of \dcp\ possible values. The neutrino mass ordering sensitivity of T2HK alone approaches $3\sigma$ C.L. However, a combined analysis with atmospheric neutrino data samples is anticipated to elevate the sensitivity to $5\sigma$ C.L or greater within a specific range of neutrino oscillation parameters. Moreover, the T2HK allows to measure $\Delta m^{2}_{32}$ and $\sin^{2}\theta_{23}$ parameters with remarkable precision, specifically 1\% for the former and 0.6 - 1.7\% for the later depending on its actual value. For simulation purpose, experimental specifications of T2HK are based on the performance of its forerunner, T2K, as detailed in Ref.~\cite{Cao:2020ans}. Tuning has been conducted to ensure compatibility with the proposed technical design report for T2HK~\cite{protocollaboration2018hyperkamiokande}, as illustrated in Appendix~\ref{app:t2hk}. 

Similar to Hyper-Kamiokande, the Deep Underground Neutrino Experiment (DUNE)~\cite{DUNE:2020lwj} is a large-scale, multipurpose neutrino experiment hosted by Fermilab, with its far detector located 1284.9~km away from the accelerator-based neutrino beam source. DUNE's far detector consists of four modular liquid argon time-projection chambers (TPC), totalling 40 kton. Over a planned thirteen-years operation (6.5 years for each neutrino and anti-neutrino modes), DUNE's far detector anticipates observing approximately 1000~$\nu_{e} (\overline{\nu}_{e})$ and 10000~$\nu_{\mu} (\overline{\nu}_{\mu})$ events~\cite{DUNE:2016evb}. This extensive statistics over such long experimental baseline enables DUNE to ascertain the neutrino mass ordering with significance of $5\sigma$ or above for all $\delta_{CP}$ actual values. In addition, DUNE can exclude CP conserving values of \dcp\ at 5$\sigma$ (3$\sigma$) or greater significance for 50\% (75\%) of the $\delta_{CP}$ true values. DUNE will improve the precision of other oscillation parameters, including $\sin^{2}\theta_{23}, \sin^{2}2\theta_{13}$. For the simulation of DUNE detector, we adopt the experimental configuration described in Ref.~\cite{DUNE:2020lwj}. 

Main experimental specifications of T2HK and DUNE, especially for measurement of neutrino oscillation with the accelerator-based neutrino sources, are highlighted in Table~\ref{tab:t2hkdune}. The complementary between the two experiments for measuring oscillation parameters and testing leptonic mixing models has been investigated extensively ~\cite{Fukasawa:2016yue,Agarwalla:2017wct}.

\bgroup
\def\arraystretch{1.2}%
\begin{table}
    \caption{\label{tab:t2hkdune}T2HK and DUNE main specifications for simulation, especially relevant to accelerator-based neutrino sources}
   \begin{tabular}{l|c|c}    Characteristics & T2HK~\cite{protocollaboration2018hyperkamiokande}   & DUNE~\cite{DUNE:2020lwj} \\ \hline
    Baseline & 295~km & 1284.9~km \\ 
    Detector type & Water Cherenkov & Liquid Argon TPC \\ 
    Detector mass& 260~kton & 40~kton \\
    Beam power & 1.3~MW & 1.2~MW\\ 
    Operation start\footnote{based on status report at NuFACT2024 conference} & 2027 & 2031 \\
    Running time $\nu:\overline{\nu}$& 2.5 yr : 7.5 yr & 6.5 yr : 6.5 yr\\
    $\nu$ energy range\footnote{relevant for oscillation measurements driven by $\Delta m^2_{31}$} & 0.1 - 1.3 GeV & 0.5 - 4.0 GeV\\
      Systematic error\footnote{the values depend on the selected samples} & signal: 3.2 - 3.9\% & signal: 2.0 - 5.0\% \\
      & background: 10\% & background: 5-20\% \\
    \hline
  \end{tabular}
\end{table}
\egroup
\subsection*{Proposed ESSnuSB and Neutrino Factory}
The European Spallation Source Neutrino Super Beam (ESSnuSB)~\cite{RosauroAlcaraz:2022str} proposes a long-baseline neutrino oscillation experiment that employs a 5~MW proton beam and a 540~kton water Cherenkov detector. The objective is to accurately measure leptonic CP violation and explore potential new physics. The high precision of \dcp, exceeding $8^{\circ}$ for all potential values of \dcp, enables ESSnuSB to evaluate certain categories of flavor models~\cite{Blennow:2020snb}. The experiment targets to measure neutrino oscillation at energies below 1~GeV, where charged-current quasi-elastic interactions predominate. A detector with 30\% PMT optical coverage achieves a detection efficiency exceeding 85\%, while maintaining a flavor mis-identification probability below 1\%. Momentum resolution for (anti-)muons induced by $\nu_{\mu}(\overline{\nu_{\mu}})$ charged-current interactions can range from 2.5\% to 5.5\%, while electron (positron) rings, produced by $\nu_e(\overline{\nu_e})$, can range from 7.5\% to 10.0\%. In simulation, a normalization uncertainty of 5\% is used for both signal and background in the selected samples. A total operational duration of approximately 10 years is assumed, divided equally between neutrino and antineutrino modes. Two candidates for the far detector site are evaluated, located at distances of $360$~km and $540$~km from the neutrino source. Our study, presented in Appendix~\ref{app:essnusbnf}, shows that the experiment with former baseline achieves greater sensitivity to both $\cos\delta_{CP}$ and $\sin^2\theta_{23}$, thereby providing enhanced utility for testing flavor models. Thus, unless otherwie specified, a baseline of $360$~km is used for ESSnuSB in our study.

Neutrino Factory (NF)~\cite{Geer:1997iz,IDS-NF:2011swj} was proposed 25 years ago, around the discovery time of neutrino oscillation by Super-K, but before the discovery of relatively large \thetae. In this facility, the neutrino beam is produced through the decay-in-flight of high-energy (anti-)muons stored within a storage ring. The well-characterized neutrino flux, comprising two equally produced flavors (muon and electron) with opposite lepton numbers, facilitates the measurement of \dcp\ and the determination of neutrino mass ordering across a broad range of \thetae\ values. A distinctive characteristic of a neutrino factory is its capability to measure neutrino oscillations at energies on the order of tens of GeV, enabling the study of transitions to tau neutrinos and facilitating tests of the unitarity of the PMNS matrix. A long baseline enables NF to examine non-standard physics scenarios across a broad parameter space with significant matter effects. The physical potentials of NF have been recently reexamined~\cite{Bogacz:2022xsj,Denton:2024glz}, considering recent developments in the field and outlining potential directions following the T2HK and DUNE era.  In NF's simulated configuration, operations over 8 years are equally divided between $\mu^{-}-$ and $\mu^{+}-$storage modes, with an assumed number of $1.06 \times 10^{21}$ POT of 50 GeV muon decay-in-flight per year. A large detector with a fiducial mass of 50 kt, comparable to that of the Super-Kamiokande, is proposed to measure the $\nu_{e}(\overline{\nu}_{e}) \xrightarrow{} \nu_{\mu} (\overline{\nu}_{\mu})$ transitions with detection efficiencies of 45\% (35\%) and the $\nu_{\mu}(\overline{\nu}_{\mu}) \xrightarrow{} \nu_{\mu} (\overline{\nu}_{\mu})$ transitions with detection efficiencies of 90\% (90\%). A supplementary detector with a fiducial mass of 5 kt is proposed for the measurement of $\nu_{e} \xrightarrow{} \nu_{\tau}$ oscillations, with a detection efficiency established at 9.6\%. The energy resolution is established at $\sigma_{E} = 15\%$ for the former detector and $20\%$ for the latter detector. For the purpose of cost-effectiveness analysis, it is assumed that these two detectors are positioned at the same baseline $L$.  Two candidates for the experimental setup at the far detector site are proposed: 4000~km and 7500~km, corresponding to muon energies of 50~GeV and 30~GeV, respectively. Appendix~\ref{app:essnusbnf} demonstrates that the baseline of 4000~km and $E_{\mu} = 50$~GeV, utilized primarily for this study unless otherwise noted, provides superior sensitivity for resolving the $\sin^{2}\theta_{23}$ octant and achieving precision in $\cos\delta_{CP}$ compared to alternative configurations.  

\section{\label{sec:result} Precision measurement of $\theta_{23}$ angle}
We utilize the GLoBES software~\cite{Huber:2004ka,Huber:2007ji} to evaluate the sensitivity of neutrino experiments to the \thetamu\ precision. The physics potential of neutrino experiments in addressing the octant of $\theta_{23}$ is characterized by which is so-called ``octant blind region," defined as the range of \sinsqthetamu\ within which the actual octant cannot be established for a given statistical significance. The predicted octant-blind region of the future neutrino experiments are shown in Table~\ref{tab:result_wth13pre_exp}. Both T2HK and DUNE experiments would be able to resolve the octant of mixing angle \thetamu\ at 3$\sigma$ C.L. if the true value of \sinsqthetamu\ lies outside of the [0.47, 0.55] range. The proposed ESSnuSB is capable of accessing a slightly narrower region of \thetamu. In neutrino factory, it is possible to investigate a wider range of mixing angle \thetamu\ in both octants. 

\begin{figure}
    \centering
    \includegraphics[width=0.48\textwidth]{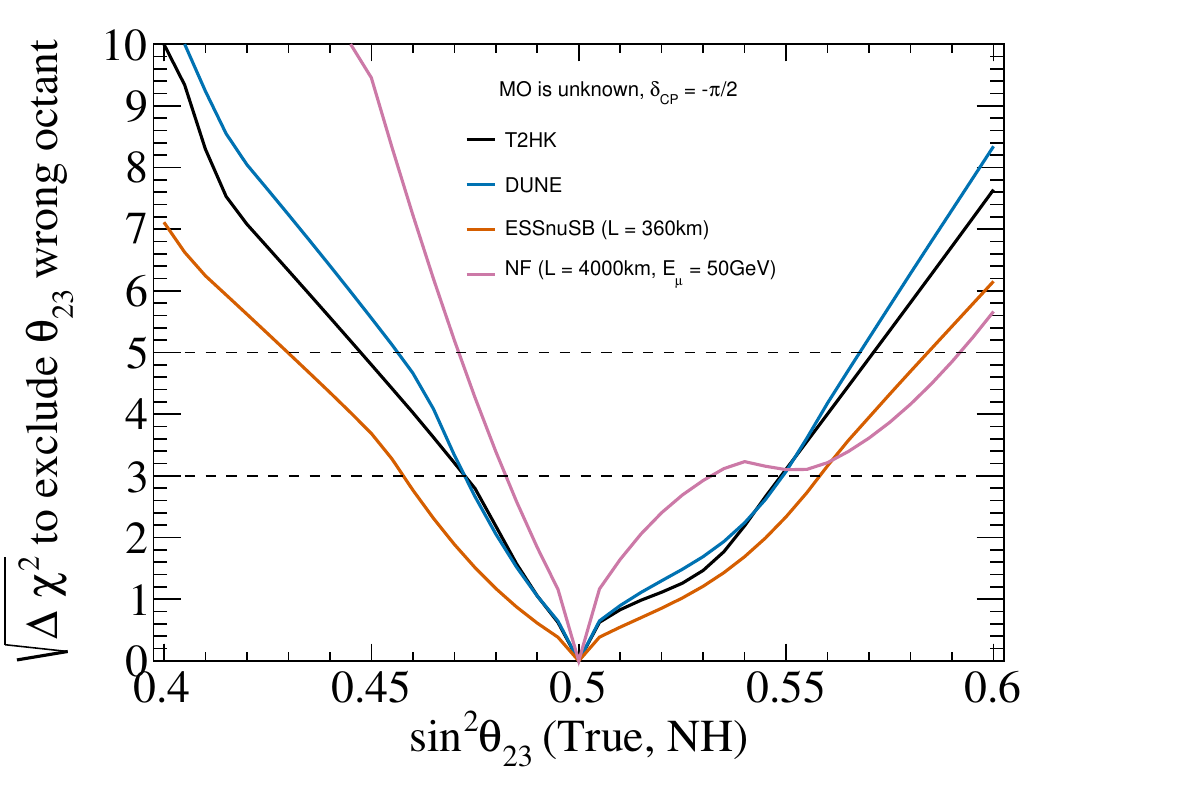}
    \caption{The statistical significance to exclude the wrong-octant as function of $\sin^{2}\theta_{23}$ with different experiments. Here $\delta_{CP} = -\pi/2$ is set, neutrino mass ordering (MO) is presumably unknown, and other relevant parameters and their uncertainties are taken from Table~\ref{tab:nuoscpara}. }
    \label{fig:octR_allexp}
\end{figure}
\begin{table}
     \centering
 \begin{tabular}{|c|wc{2cm}|wc{2cm}|}
 \hline
  Exp. & \multicolumn{2}{c|}{ Octant-blind regions of \sinsqthetamu}\\ \hline \cline{2-3}
  & 3$\sigma$ C.L.  & 5$\sigma$ C.L. \\
  \hline
   T2HK & [0.47, 0.55]  & [0.45, 0.57]   \\
 \hline
   DUNE & [0.47, 0.55]   & [0.46, 0.57]   \\
 \hline
   ESSnuSB & [0.46, 0.56] & [0.43, 0.58]  \\
   \hline
   Nu-Factory & [0.48, 0.53]  & [0.47, 0.59] \\
   \hline
   \end{tabular}
  \caption{\label{tab:result_wth13pre_exp}The octant-blind regions at 3$\sigma$ and 5$\sigma$ C.L. of $\theta_{23}$ with individual T2HK and DUNE, ESSnuSB, and Neutrino Factory. Here the current constraint of \thetae\ is used.}
\end{table}
\noindent The numbers presented in Table~\ref{tab:result_wth13pre_exp} are calculated based on the existing limitations of the leptonic mixing parameters. An estimation using a 1\% uncertainty on the \sinsqthetareac\ reveals that the 3$\sigma$ C.L. octant blind region can be narrowed down by less than 10\%. More precisely, the precentage is around 8\% for T2HK, 5\% for DUNE and 9\% for a combined analysis of T2HK and DUNE. While the precision of $\theta_{12}$ can achieve sub-percent accuracy with JUNO~\cite{JUNO:2022mxj}, its effect on the precision of $\theta_{23}$ is minimal. Nevertheless, this unprecedented $\theta_{12}$ constraint serves to narrow down the allowed range of parameters predicted by the flavor models outlined in Section~\ref{sec:flavormodel}, allowing data to differentiate the models.
 
\begin{figure*}
\includegraphics[width=0.48\textwidth]{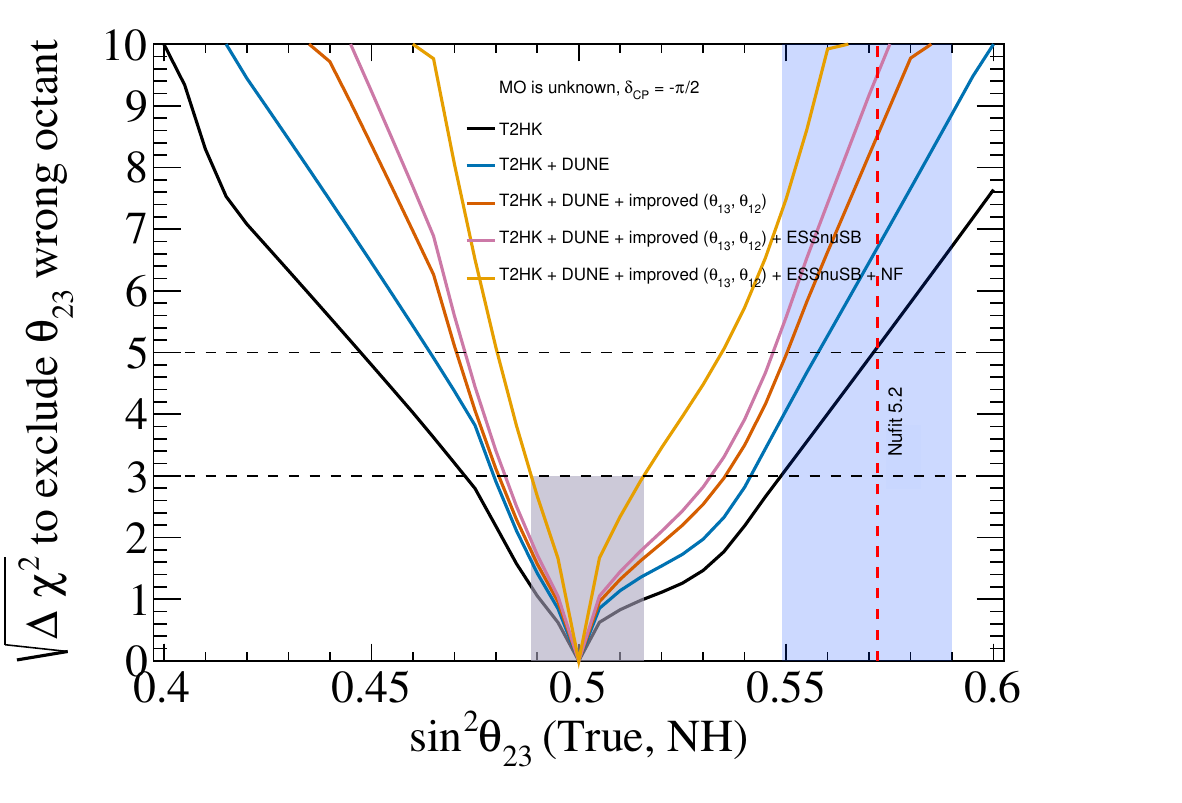}
\includegraphics[width=0.48\textwidth]{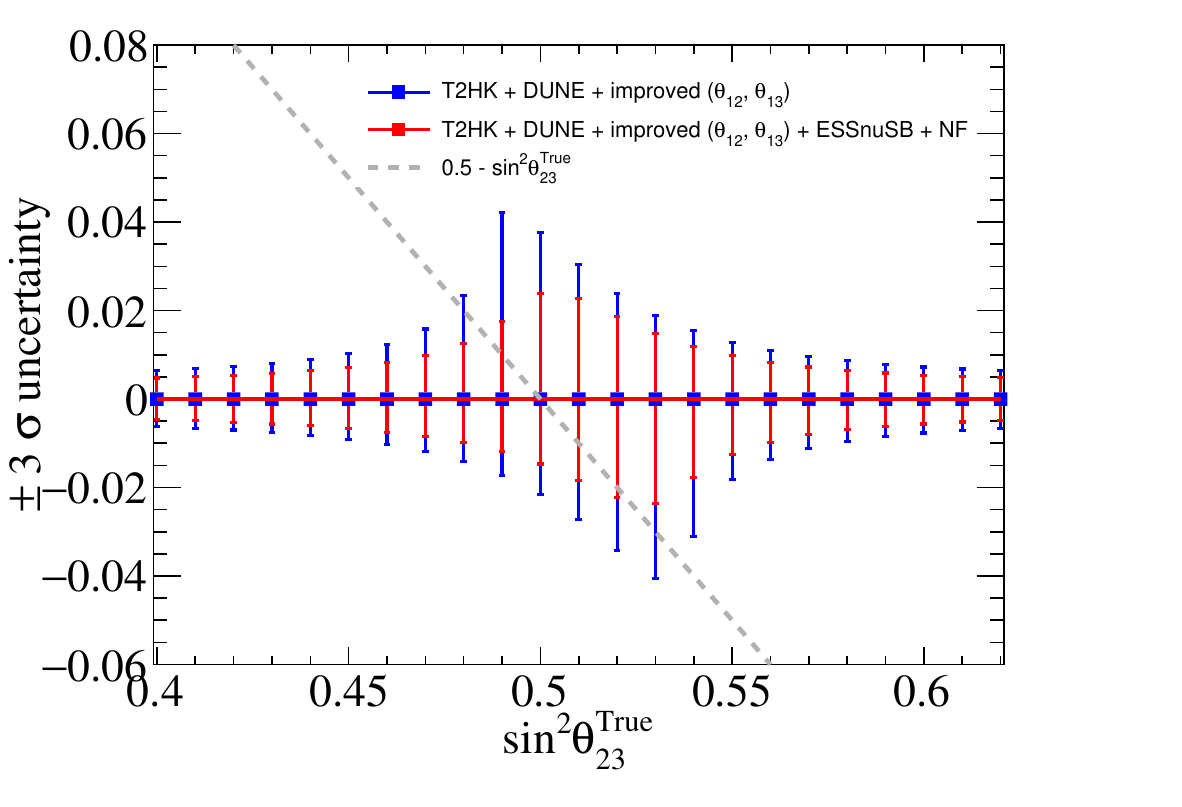}

\caption{\label{fig:ORhyperkdune}The left plot presents  statistical significance to exclude the  wrong-octant as a function of $\sin^{2}\theta_{23}$ with T2HK and various configurations of joint analyses. The right plot is the $\pm 3\sigma$ uncertainty as function of true values of \sinsqthetamu\ with joint analyses. Here $\delta_{CP} =  -\pi/2$ is set, MO is presumably unknown. }
\end{figure*}

Our simulation, depicted in Fig.~\ref{fig:ORhyperkdune}, suggests that T2HK and DUNE will conclusively establish the higher octant of \thetamu\ for all values of \thetamu\ within the currently fitted $\pm 1\sigma$ interval of parameter \sinsqthetamu\ = 0.572$^{+ 0.018}_{- 0.023}$. However, if the actual value of \sinsqthetamu\ falls within the range of $[0.48,0.54]$, its ambiguity can not be solved definitively with a statistical significance of 3$\sigma$ C.L. or higher, even with the use of T2HK, DUNE, or their joint analysis with envisaged constraints on \thetae\ and \thetasol. In this scenario, possible solution to enhancing the reach of \thetamu\ is to do a combined analysis of the T2HK and DUNE with the proposed ESSnuSB and Neutrino Factory. As shown in Fig.~\ref{fig:ORhyperkdune} and reported in Table.~\ref{tab:octant_blind_jointana}, the octant ambiguity can be resolved at 3$\sigma$ C.L. if the truth value of \sinsqthetamu\ lies outside of $[0.49, 0.52]$. This corresponds to a 50\% reduction in the octant-blind region that remains inaccessible through a joint analysis of T2HK and DUNE. The interval $[0.49, 0.52]$ is anticipated as the ``dead" region for \thetamu\ octant measurement. 


\begin{table}
     \centering
\begin{tabular}{|c|wc{2cm}|wc{2cm}|}
 \hline
 Joint analysis & \multicolumn{2}{c|}{Octant-blind region of $\sin^{2}\theta_{23}$}\\ 
 \cline{2-3}  & $3\sigma$ C.L. & $5\sigma$ C.L. \\
 \hline
 T2HK & [0.47, 0.55] &  [0.45, 0.57] \\
 + DUNE & [0.48, 0.54]   & [0.46, 0.56] \\
 + improved ($\theta_{13}, \theta_{12})$ &  [0.48, 0.54] & [0.47, 0.55]\\
 + ESSnuSB & [0.48, 0.53]& [0.47, 0.55] \\
 + Neutrino Factory & [0.49, 0.52] & [0.48, 0.53]  \\
 \hline
   \end{tabular}
\caption{\label{tab:octant_blind_jointana}Octant-blind regions at  3$\sigma$ and 5$\sigma$ C.L. of $\theta_{23}$ with staging joint analyses }
\end{table}

\section{\label{sec:flavormodel} Implications for the leptonic flavor models}
\begin{figure}
    \centering
    \includegraphics[width=0.48\textwidth]{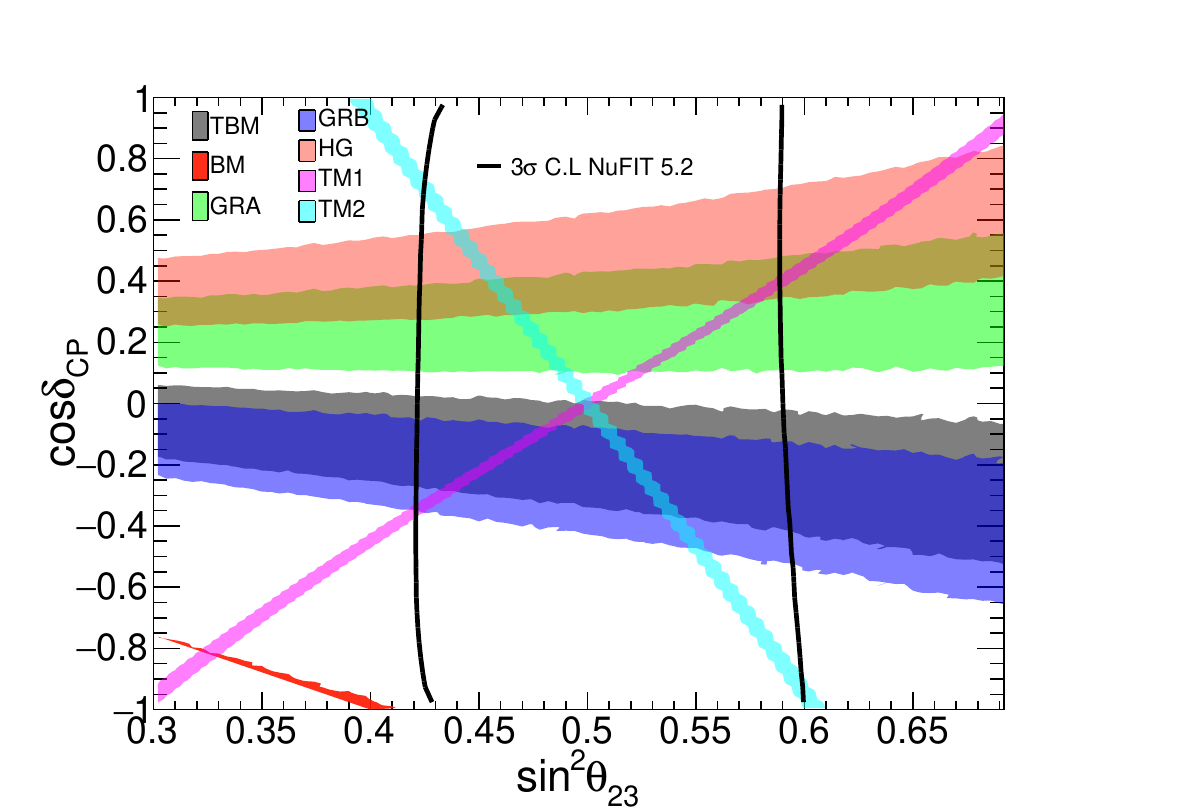}
    \caption{Allowed parameter spaces of (\dcp,\thetamu) with flavor models are overlaid with current constraint of the global neutrino data.}
    \label{fig:flavmodelnow}
\end{figure}

\begin{figure}
    \centering
    \includegraphics[width=0.48\textwidth]{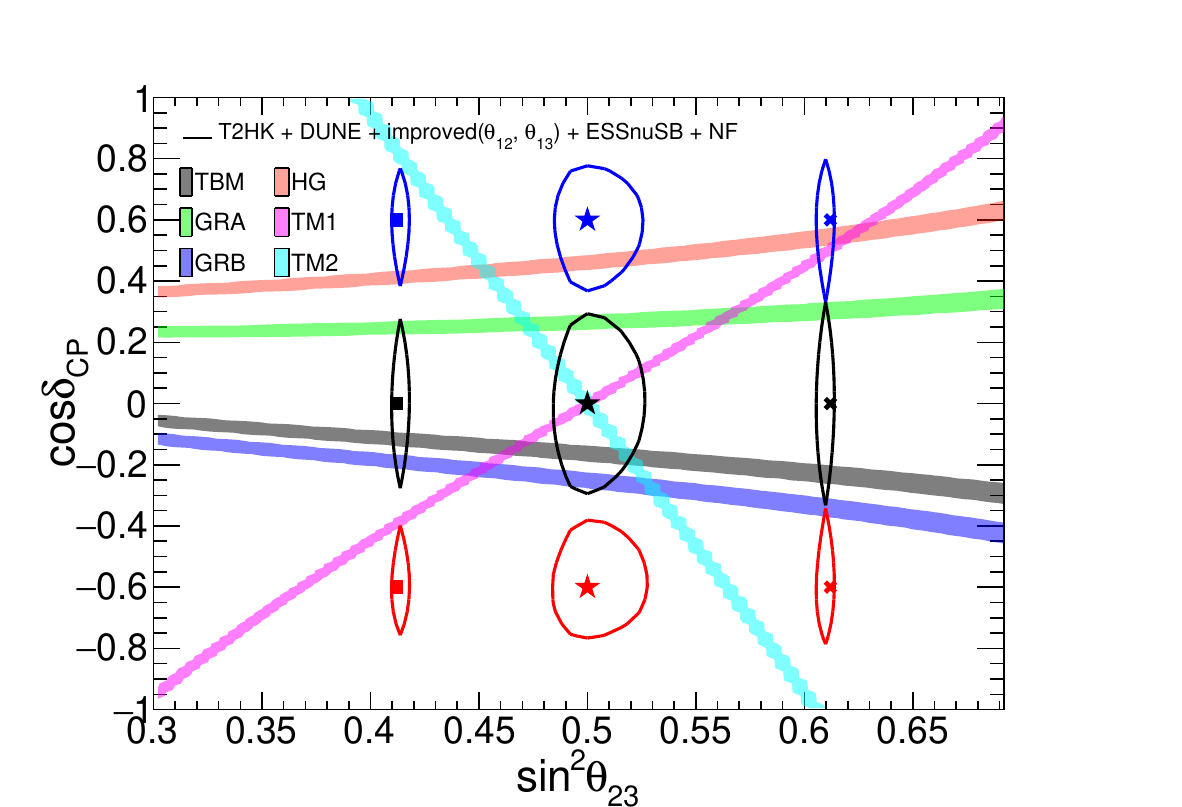}
    \caption{The allowed parameter space of (\dcp,\thetamu) with flavor models is tightened when applying more precise measurements of \thetae, \thetasol. The sensitivity contours of future constraints at some actual values of parameters are  computed with joint analysis of simulated data using T2HK, DUNE, ESSnuSB and Neutrino Factory.}
    \label{fig:flavmodelfut}
\end{figure}

 That leptonic mixing, represented by two relatively large angles $\theta_{23}\approx \pi/4$, $\theta_{12}\approx \pi/6$, and one small angle $\theta_{13}\approx \pi/20$ in the PMNS matrix, is so different from the quark mixing, expressed by approximate unity CKM matrix, triggers questions whether the underlying symmetry is responsible for such patterns and how to describe them in a unified framework. Apparently, the question is relevant to unknown nature of neutrino mass. In this analysis, we examine a category of theoretical models in which the leptonic mixing matrix arises from the misalignment between the \emph{flavor-definite} states $\{l_{\alpha},\nu_{\alpha} \}$ and \emph{symmetry} states $\{\Tilde{l}_{\alpha},\Tilde{\nu}_{\alpha} \}$ of both charged and neutral leptons~\cite{Mohapatra:2006gs}. The weak charged-current is written as $J_{\mu} = \overline{\Tilde{l}}\gamma^{\mu}(1-\gamma_5)\Tilde{\nu}\equiv \overline{l}\gamma^{\mu}(1-\gamma_5)U^{\dagger}_lU_{\nu}\nu$ resulting in a formulation $U_{\text{PMNS}} = U^{\dagger}_lU_{\nu}$ where $U_l$ and $U_{\nu}$ are unitary transformation matrices which rotate the states: $\Tilde{l} = U_l l $ and $\Tilde{\nu} = U_{\nu} \nu $; and diagonalize the mass matrices $U^{\dagger}_l M_lM_l^{\dagger}U_l = \text{diag}(m_e^2,m_{\mu}^2, m_{\tau}^2)$ and $U_{\nu}^TM_{\nu}U_{\nu}=\text{diag}(m_1,m_2,m_3)$. The mass matrices of charged and neutral leptons are assumed to be invariant when subjected to non-Abelian discrete symmetries $G_l$ and $G_{\nu}$ respectively, which were commonly originated from a family symmetry $G_f$, see ~\cite{Altarelli:2010gt,King:2013eh} and references therein. Several $\{G_f\rightarrow G_l\oplus G_{\nu}\}$ models have been proposed to examine whether the hypothetical patterns of the leptonic mixing agrees with the measured neutrino parameters or not.  Among them, the class of flavor models that involve the predictability of the less known \dcp\ based on its relationship with other more precisely constrained mixing angles is of noteworthy interest because of its testability. The so-called \emph{solar} sum rules~\cite{Petcov:2014laa}, in which $U_{\nu}$ is characterized by a maximal $\theta_{23}^{\nu}=\pi/4$ rotation in 2-3 generation plane, a predetermined $\theta_{12}^{\nu}$ in 1-2 generation plane, and non-diagonal $U_l$ to account for relatively smallness of the \thetae\ angle, establish following relation:
 \begin{align} \label{eq:solar_sum_rule}
    \cos\delta_{CP} &= \frac{\tan\theta_{23}}{\sin2\theta_{12}\sin\theta_{13}}\left[\cos2\theta_{12}^{\nu} \right. \\ \nonumber
    &+ \left. (\sin^{2}\theta_{12} - \cos^{2}\theta_{12}^{\nu})(1 - \cot^{2}\theta_{23}\sin^{2}\theta_{13})\right].
\end{align}
The values of $\sin \theta_{12}^{\nu}$ are fixed at (i) $1/\sqrt{2}$, (ii) $ 1/\sqrt{3}$, (iii) 1/2, (iv)$1/\sqrt{\sqrt{5}r_g}$, (v) $\sqrt{3-r_g}/2$, (vi)$\sqrt{(2-r_g)/3}$ where golden ratio $r_g = (1 + \sqrt{5})/2$, depending on the symmetry form of the $U_{\nu}$ matrices, which are, respectively, (i) bi-maximal (BM), (ii) tri-bimaximal  (TBM), (iii) hexagonal (HG), (iv) golden rule type A (GRA), (v) golden rule type B (GRB), and (vi) golden rule type C (GRC). Eq.~(\ref{eq:solar_sum_rule}) exhibits a non-negligible dependence of \dcp\ on the \thetamu, particularly deviation of $\cos\delta_{CP}$ from zero is greater as the value of \thetamu\ increases. It is observed that the BM and GRC models are disfavored since a valid \dcp\ acquires \thetamu\ to be out of the current 3$\sigma$ C.L. allowed range of this parameter. The remaining four models predict significant amount of the CP violation. Alternative models, in which $U_l$ is assumed to be diagonal and symmetry pattern of $U_{\nu}$ to be partially broken to formulate empirically the observed PMNS matrix, have also been investigated~\cite{Costa:2023bxw}. There are two \emph{atmospheric} sum rules, known as TM1 and TM2, depending on the first or second column of the tri-bimaximal form of $U_{\nu}$ kept to be conserved. TM1 and TM2 establish a relation between \thetasol\ and \thetae\ mixing angles, specified as $\sin^2\theta_{12}=\frac{1-3\sin^2\theta_{13}}{3(1-\sin^2\theta_{13})}$ and $\sin^2\theta_{12}=\frac{1}{3(1-\sin^2\theta_{13})}$ respectively. The predictions are consistent in 3$\sigma$ C.L. with the current constraint of oscillation parameters shown in Table~\ref{tab:nuoscpara}. These predictions will be further tested by future neutrino experiments, when the precision of both \thetasol\ and \thetae\ mixing angle will be improved.  Moreover, TM1 and TM2 enforce a relationship between \dcp\ and other mixing parameters~\cite{Costa:2023bxw}, provided in Eq.~(\ref{eq:atm_sumrule}). 
\begin{align}\label{eq:atm_sumrule}
     \cos\delta_{CP}= 
\begin{cases}
   -\frac{(1 - 5\sin^{2}\theta_{13})\cot2\theta_{23}}{2\sqrt{2}\sin\theta_{13}\sqrt{1 - 3\sin^{2}\theta_{13}}}\ (\text{TM1})\\
    \frac{(1-2\sin^2\theta_{13})\cot2\theta_{23}}{\sin\theta_{13}\sqrt{2 - 3\sin^{2}\theta_{13}}}\ (\text{TM2})   
\end{cases}
\end{align}
In comparison to the \emph{solar} sum rules, the \emph{atmospheric} sum rules exhibit a more profound correlation between the value of \dcp\ and the value of \thetamu. The allowed parameter space of (\dcp, \thetamu) of the above-mentioned flavor models are presented in Fig.~\ref{fig:flavmodelnow}. Given the current 3$\sigma$ C.L. range of \sinsqthetamu, all models, with the exception of the BM and GRC which are not consistent with data, predict a substantial violation of the CP symmetry in the lepton sector. The difference between \emph{solar} sum rules are relatively small due to the uncertainties on \thetasol\ and \thetae.
\noindent Fig.~\ref{fig:flavmodelfut} shows the allowed parameter space in (\dcp, \thetamu) of the flavor models, assuming that the central values of $\sin^2\theta_{12}$ and $\sin^2\theta_{13}$ remain unchanged, but their uncertainty is decreased to 0.5\% and 1.0\% respectively. In comparison to Fig.~\ref{fig:flavmodelnow}, the predicted parameter space from different \emph{solar} sum rules become more distinguishable, thereby enhancing the test capability. We find that if CP is conserved, a joint analysis of T2HK and DUNE would be sensitive enough to eliminate all \emph{solar} and \emph{atmospheric} TM1 sum rules for all possible allowed value of \thetamu. The CP conservation is still applicable for TM2 if \thetamu\ is significantly deviated from the maximal mixing, specifically when actual value of \sinsqthetamu\ is close to 0.4 in lower octant or 0.6 in higher octant. Thus the measurement of \thetamu\ (in addition to \thetasol) measurement will settle the test for TM2. For maximum CP violation ($\cos\delta_{\text{CP}}=0$), we find that uncertainty of $\cos\delta_{\text{CP}}$ of the combination between T2HK and DUNE, which is around the range of [-0.4, 0.5] at 3$\sigma$ C.L., is not sufficient for differentiating the \emph{solar} and \emph{atmospheric} TM1 sum rule. If actual \sinsqthetamu\  falls outside of [0.44, 0.56], the analysis of T2HK and DUNE will rule out the \emph{atmospheric} TM2 sum rule. Fig.~\ref{fig:flavmodelfut} includes a possibility of $\cos\delta_{CP} - \sin^{2}\theta_{23}$  sensitivity using a combined T2HK, DUNE, ESSnuSB and Neutrino Factory. It is found that this ultimate analysis, which can achieve precision of $\cos\delta_{\text{CP}}$ around the range of [-0.25, 0.30], will rule out the \emph{solar} HG sum rule in entire parameter space of \thetamu. Also, the enhancement in the $\cos\delta_{\text{CP}}$ precision also helps to eliminate the \emph{atmospheric} TM1 and TM2 models if $\sin^{2}\theta_{23}$ is found to out of the range of [0.44, 0.57] and [0.47, 0.53] respectively. However, the sensitivity of this ultimate reach on $\cos\delta_{CP}$ is not sufficient to distinguish the \emph{solar} TBM, GRA, GRB sum rules.

\section{\label{sec:summarize} Summary}
The present landscape of the neutrino oscillation measurement indicates that \thetamu\ is in close proximity to the maximal mixing. We explore the optimum extent of this mixing angle by analysis of forthcoming influential experiments, T2HK and DUNE with potential future inclusion of ESSnuSB and the neutrino factory. Our findings indicate that the accuracy of this measurement does not depend significantly on the truth value as well as the precision of \dcp. Furthermore, it is not affected by the unidentified neutrino mass ordering. As long as the actual value falls within the current $\pm 1\sigma$ interval, both T2HK and DUNE can assert the higher octant of \thetamu. However, if the value of \thetamu\ is within the octant-blind region of [0.48, 0.54] at $3\sigma $ C.L., these experiments cannot resolve the ambiguity of this parameter. A possible solution is to have a joint analysis of T2HK and DUNE with ESSnuSB and Neutrino Factory, provided that the data is available. Within this scenario, the octant-blind region will be narrowed down to a range of [0.49, 0.52]. The precision measurement of \thetamu, together with the \dcp\ precision, are essential for testing a category of flavor models that result in an exact relationship between these two parameters via the \emph{solar} and \emph{atmospheric} sum rules. We find that if CP is conserved, the combined sensitivity between T2HK and DUNE  will eliminate nearly all of these models. In the case that the CP is maximally violated, only \emph{atmospheric} TM2 can be excluded if real value of \thetamu\ is significantly deviated from $\pi/4$. Conducting a combined analysis of the two aforementioned experiments with ESSnuSB and Neutrino factory, if available, can effectively eliminate the \emph{solar} HG sum rules and both \emph{atmospheric} TM1 and TM2 sum rules in certain range of \thetamu. Nevertheless, the overall sensitivity is inadequate to tell preference of data to particular \emph{solar} sum rule model among TBM, GRA, and GRB models.

\section*{Acknowledgement}
S. Cao would like to thank KEK for their hospitality during his visit. A. Nath wishes to express gratitude to IFIRSE, ICISE for their hospitality.  Phan To Quyen was funded by the Master, PhD Scholarship Programme of Vingroup Innovation Foundation (VINIF), code VINIF.2023.TS.095. The research of S. Cao and N. T. H. Van is funded by the National Foundation for Science and Technology Development (NAFOSTED) of Vietnam under Grant No. 103.99-2023.144. 

\bibliography{Reference}

\begin{thebibliography}{43}%
\makeatletter
\providecommand \@ifxundefined [1]{%
 \@ifx{#1\undefined}
}%
\providecommand \@ifnum [1]{%
 \ifnum #1\expandafter \@firstoftwo
 \else \expandafter \@secondoftwo
 \fi
}%
\providecommand \@ifx [1]{%
 \ifx #1\expandafter \@firstoftwo
 \else \expandafter \@secondoftwo
 \fi
}%
\providecommand \natexlab [1]{#1}%
\providecommand \enquote  [1]{``#1''}%
\providecommand \bibnamefont  [1]{#1}%
\providecommand \bibfnamefont [1]{#1}%
\providecommand \citenamefont [1]{#1}%
\providecommand \href@noop [0]{\@secondoftwo}%
\providecommand \href [0]{\begingroup \@sanitize@url \@href}%
\providecommand \@href[1]{\@@startlink{#1}\@@href}%
\providecommand \@@href[1]{\endgroup#1\@@endlink}%
\providecommand \@sanitize@url [0]{\catcode `\\12\catcode `\$12\catcode
  `\&12\catcode `\#12\catcode `\^12\catcode `\_12\catcode `\%12\relax}%
\providecommand \@@startlink[1]{}%
\providecommand \@@endlink[0]{}%
\providecommand \url  [0]{\begingroup\@sanitize@url \@url }%
\providecommand \@url [1]{\endgroup\@href {#1}{\urlprefix }}%
\providecommand \urlprefix  [0]{URL }%
\providecommand \Eprint [0]{\href }%
\providecommand \doibase [0]{http://dx.doi.org/}%
\providecommand \selectlanguage [0]{\@gobble}%
\providecommand \bibinfo  [0]{\@secondoftwo}%
\providecommand \bibfield  [0]{\@secondoftwo}%
\providecommand \translation [1]{[#1]}%
\providecommand \BibitemOpen [0]{}%
\providecommand \bibitemStop [0]{}%
\providecommand \bibitemNoStop [0]{.\EOS\space}%
\providecommand \EOS [0]{\spacefactor3000\relax}%
\providecommand \BibitemShut  [1]{\csname bibitem#1\endcsname}%
\let\auto@bib@innerbib\@empty
\bibitem [{\citenamefont {Fukuda}\ \emph {et~al.}(1998)\citenamefont {Fukuda}
  \emph {et~al.}}]{fukuda1998evidence}%
  \BibitemOpen
  \bibfield  {author} {\bibinfo {author} {\bibfnamefont {Y.}~\bibnamefont
  {Fukuda}} \emph {et~al.} (\bibinfo {collaboration} {Super-Kamiokande
  Collaboration}),\ }\bibfield  {title} {\enquote {\bibinfo {title} {{Evidence
  for oscillation of atmospheric neutrinos}},}\ }\href {\doibase
  10.1103/PhysRevLett.81.1562} {\bibfield  {journal} {\bibinfo  {journal}
  {Phys. Rev. Lett.}\ }\textbf {\bibinfo {volume} {81}},\ \bibinfo {pages}
  {1562} (\bibinfo {year} {1998})},\ \Eprint
  {http://arxiv.org/abs/hep-ex/9807003}{arXiv:hep-ex/9807003}\BibitemShut
  {NoStop}%
\bibitem [{\citenamefont {Ahmad}\ \emph {et~al.}(2001)\citenamefont {Ahmad}
  \emph {et~al.}}]{Ahmad:2001an}%
  \BibitemOpen
  \bibfield  {author} {\bibinfo {author} {\bibfnamefont {Q.}~\bibnamefont
  {Ahmad}} \emph {et~al.} (\bibinfo {collaboration} {SNO Collaboration}),\
  }\bibfield  {title} {\enquote {\bibinfo {title} {Measurement of the rate of
  $\nu_{e}+d \to p+p+e^-$ interactions produced by {$^8B$} solar neutrinos at
  the {S}udbury {N}eutrino {O}bservatory},}\ }\href {\doibase
  10.1103/PhysRevLett.87.071301} {\bibfield  {journal} {\bibinfo  {journal}
  {Phys. Rev. Lett.}\ }\textbf {\bibinfo {volume} {87}},\ \bibinfo {pages}
  {071301} (\bibinfo {year} {2001})},\ \Eprint
  {http://arxiv.org/abs/0106015}{arXiv:0106015 [nucl-ex]}\BibitemShut {NoStop}%
\bibitem [{\citenamefont {Ahmad}\ \emph {et~al.}(2002)\citenamefont {Ahmad}
  \emph {et~al.}}]{ahmad2002measurement}%
  \BibitemOpen
  \bibfield  {author} {\bibinfo {author} {\bibfnamefont {Q.~R.}\ \bibnamefont
  {Ahmad}} \emph {et~al.} (\bibinfo {collaboration} {SNO}),\ }\bibfield
  {title} {\enquote {\bibinfo {title} {{Measurement of day and night neutrino
  energy spectra at SNO and constraints on neutrino mixing parameters}},}\
  }\href {\doibase 10.1103/PhysRevLett.89.011302} {\bibfield  {journal}
  {\bibinfo  {journal} {Phys. Rev. Lett.}\ }\textbf {\bibinfo {volume} {89}},\
  \bibinfo {pages} {011302} (\bibinfo {year} {2002})},\ \Eprint
  {http://arxiv.org/abs/nucl-ex/0204009}{arXiv:nucl-ex/0204009}\BibitemShut
  {NoStop}%
\bibitem [{\citenamefont {Navas}\ \emph {et~al.}(2024)\citenamefont {Navas}
  \emph {et~al.}}]{ParticleDataGroup:2024cfk}%
  \BibitemOpen
  \bibfield  {author} {\bibinfo {author} {\bibfnamefont {S.}~\bibnamefont
  {Navas}} \emph {et~al.} (\bibinfo {collaboration} {Particle Data Group}),\
  }\bibfield  {title} {\enquote {\bibinfo {title} {{Review of particle
  physics}},}\ }\href {\doibase 10.1103/PhysRevD.110.030001} {\bibfield
  {journal} {\bibinfo  {journal} {Phys. Rev. D}\ }\textbf {\bibinfo {volume}
  {110}},\ \bibinfo {pages} {030001} (\bibinfo {year} {2024})}\BibitemShut
  {NoStop}%
\bibitem [{\citenamefont {Pontecorvo}(1968)}]{pontecorvo1968neutrino}%
  \BibitemOpen
  \bibfield  {author} {\bibinfo {author} {\bibfnamefont {B.}~\bibnamefont
  {Pontecorvo}},\ }\bibfield  {title} {\enquote {\bibinfo {title} {Neutrino
  experiments and the problem of conservation of leptonic charge},}\
  }\href@noop {} {\bibfield  {journal} {\bibinfo  {journal} {Sov. Phys. JETP}\
  }\textbf {\bibinfo {volume} {26}},\ \bibinfo {pages} {165} (\bibinfo {year}
  {1968})}\BibitemShut {NoStop}%
\bibitem [{\citenamefont {Maki}\ \emph {et~al.}(1962)\citenamefont {Maki},
  \citenamefont {Nakagawa},\ and\ \citenamefont {Sakata}}]{maki1962remarks}%
  \BibitemOpen
  \bibfield  {author} {\bibinfo {author} {\bibfnamefont {Z.}~\bibnamefont
  {Maki}}, \bibinfo {author} {\bibfnamefont {M.}~\bibnamefont {Nakagawa}}, \
  and\ \bibinfo {author} {\bibfnamefont {S.}~\bibnamefont {Sakata}},\
  }\bibfield  {title} {\enquote {\bibinfo {title} {Remarks on the unified model
  of elementary particles},}\ }\href {\doibase 10.1143/PTP.28.870} {\bibfield
  {journal} {\bibinfo  {journal} {Prog. Theor. Phys.}\ }\textbf {\bibinfo
  {volume} {28}},\ \bibinfo {pages} {870} (\bibinfo {year} {1962})}\BibitemShut
  {NoStop}%
\bibitem [{\citenamefont {Abe}\ \emph {et~al.}(2023)\citenamefont {Abe} \emph
  {et~al.}}]{T2K:2023smv}%
  \BibitemOpen
  \bibfield  {author} {\bibinfo {author} {\bibfnamefont {K.}~\bibnamefont
  {Abe}} \emph {et~al.} (\bibinfo {collaboration} {T2K}),\ }\bibfield  {title}
  {\enquote {\bibinfo {title} {{Measurements of neutrino oscillation parameters
  from the T2K experiment using $3.6\times 10^{21}$ protons on target}},}\
  }\href {\doibase 10.1140/epjc/s10052-023-11819-x} {\bibfield  {journal}
  {\bibinfo  {journal} {Eur. Phys. J. C}\ }\textbf {\bibinfo {volume} {83}},\
  \bibinfo {pages} {782} (\bibinfo {year} {2023})},\ \Eprint
  {http://arxiv.org/abs/2303.03222}{arXiv:2303.03222 [hep-ex]}\BibitemShut
  {NoStop}%
\bibitem [{\citenamefont {Acero}\ \emph
  {et~al.}(2022{\natexlab{a}})\citenamefont {Acero}, \citenamefont {Adamson},
  \citenamefont {Aliaga}, \citenamefont {Anfimov}, \citenamefont {Antoshkin},
  \citenamefont {Arrieta-Diaz}, \citenamefont {Asquith}, \citenamefont
  {Aurisano}, \citenamefont {Back}, \citenamefont {Backhouse} \emph
  {et~al.}}]{acero2022improved}%
  \BibitemOpen
  \bibfield  {author} {\bibinfo {author} {\bibfnamefont {M.}~\bibnamefont
  {Acero}}, \bibinfo {author} {\bibfnamefont {P.}~\bibnamefont {Adamson}},
  \bibinfo {author} {\bibfnamefont {L.}~\bibnamefont {Aliaga}}, \bibinfo
  {author} {\bibfnamefont {N.}~\bibnamefont {Anfimov}}, \bibinfo {author}
  {\bibfnamefont {A.}~\bibnamefont {Antoshkin}}, \bibinfo {author}
  {\bibfnamefont {E.}~\bibnamefont {Arrieta-Diaz}}, \bibinfo {author}
  {\bibfnamefont {L.}~\bibnamefont {Asquith}}, \bibinfo {author} {\bibfnamefont
  {A.}~\bibnamefont {Aurisano}}, \bibinfo {author} {\bibfnamefont
  {A.}~\bibnamefont {Back}}, \bibinfo {author} {\bibfnamefont {C.}~\bibnamefont
  {Backhouse}},  \emph {et~al.},\ }\bibfield  {title} {\enquote {\bibinfo
  {title} {Improved measurement of neutrino oscillation parameters by the
  no$\nu$a experiment},}\ }\href@noop {} {\bibfield  {journal} {\bibinfo
  {journal} {Physical Review D}\ }\textbf {\bibinfo {volume} {106}},\ \bibinfo
  {pages} {032004} (\bibinfo {year} {2022}{\natexlab{a}})}\BibitemShut
  {NoStop}%
\bibitem [{\citenamefont {Wester}\ \emph {et~al.}(2024)\citenamefont {Wester}
  \emph {et~al.}}]{Super-Kamiokande:2023ahc}%
  \BibitemOpen
  \bibfield  {author} {\bibinfo {author} {\bibfnamefont {T.}~\bibnamefont
  {Wester}} \emph {et~al.} (\bibinfo {collaboration} {Super-Kamiokande}),\
  }\bibfield  {title} {\enquote {\bibinfo {title} {{Atmospheric neutrino
  oscillation analysis with neutron tagging and an expanded fiducial volume in
  Super-Kamiokande I\textendash{}V}},}\ }\href {\doibase
  10.1103/PhysRevD.109.072014} {\bibfield  {journal} {\bibinfo  {journal}
  {Phys. Rev. D}\ }\textbf {\bibinfo {volume} {109}},\ \bibinfo {pages}
  {072014} (\bibinfo {year} {2024})},\ \Eprint
  {http://arxiv.org/abs/2311.05105}{arXiv:2311.05105 [hep-ex]}\BibitemShut
  {NoStop}%
\bibitem [{\citenamefont {Djurcic}\ \emph {et~al.}(2015)\citenamefont {Djurcic}
  \emph {et~al.}}]{djurcic2015juno}%
  \BibitemOpen
  \bibfield  {author} {\bibinfo {author} {\bibfnamefont {Z.}~\bibnamefont
  {Djurcic}} \emph {et~al.} (\bibinfo {collaboration} {JUNO Collaboration}),\
  }\bibfield  {title} {\enquote {\bibinfo {title} {{JUNO Conceptual Design
  Report}},}\ }\href@noop {} {\  (\bibinfo {year} {2015})},\ \Eprint
  {http://arxiv.org/abs/1508.07166}{arXiv:1508.07166
  [physics.ins-det]}\BibitemShut {NoStop}%
\bibitem [{\citenamefont {Cao}\ \emph {et~al.}(2021)\citenamefont {Cao},
  \citenamefont {Nath}, \citenamefont {Ngoc}, \citenamefont {Quyen},
  \citenamefont {Hong~Van},\ and\ \citenamefont {Francis}}]{Cao:2020ans}%
  \BibitemOpen
  \bibfield  {author} {\bibinfo {author} {\bibfnamefont {S.}~\bibnamefont
  {Cao}}, \bibinfo {author} {\bibfnamefont {A.}~\bibnamefont {Nath}}, \bibinfo
  {author} {\bibfnamefont {T.~V.}\ \bibnamefont {Ngoc}}, \bibinfo {author}
  {\bibfnamefont {P.~T.}\ \bibnamefont {Quyen}}, \bibinfo {author}
  {\bibfnamefont {N.~T.}\ \bibnamefont {Hong~Van}}, \ and\ \bibinfo {author}
  {\bibfnamefont {N.~K.}\ \bibnamefont {Francis}},\ }\bibfield  {title}
  {\enquote {\bibinfo {title} {{Physics potential of the combined sensitivity
  of T2K-II, NO$\nu$A extension, and JUNO}},}\ }\href {\doibase
  10.1103/PhysRevD.103.112010} {\bibfield  {journal} {\bibinfo  {journal}
  {Phys. Rev. D}\ }\textbf {\bibinfo {volume} {103}},\ \bibinfo {pages}
  {112010} (\bibinfo {year} {2021})},\ \Eprint
  {http://arxiv.org/abs/2009.08585}{arXiv:2009.08585 [hep-ph]}\BibitemShut
  {NoStop}%
\bibitem [{\citenamefont {Cabrera}\ \emph {et~al.}(2022)\citenamefont {Cabrera}
  \emph {et~al.}}]{Cabrera:2020ksc}%
  \BibitemOpen
  \bibfield  {author} {\bibinfo {author} {\bibfnamefont {A.}~\bibnamefont
  {Cabrera}} \emph {et~al.},\ }\bibfield  {title} {\enquote {\bibinfo {title}
  {{Synergies and prospects for early resolution of the neutrino mass
  ordering}},}\ }\href {\doibase 10.1038/s41598-022-09111-1} {\bibfield
  {journal} {\bibinfo  {journal} {Sci. Rep.}\ }\textbf {\bibinfo {volume}
  {12}},\ \bibinfo {pages} {5393} (\bibinfo {year} {2022})},\ \Eprint
  {http://arxiv.org/abs/2008.11280}{arXiv:2008.11280 [hep-ph]}\BibitemShut
  {NoStop}%
\bibitem [{\citenamefont {Abe}\ \emph {et~al.}(2020)\citenamefont {Abe} \emph
  {et~al.}}]{T2K:2019bcf}%
  \BibitemOpen
  \bibfield  {author} {\bibinfo {author} {\bibfnamefont {K.}~\bibnamefont
  {Abe}} \emph {et~al.} (\bibinfo {collaboration} {T2K}),\ }\bibfield  {title}
  {\enquote {\bibinfo {title} {{Constraint on the matter\textendash{}antimatter
  symmetry-violating phase in neutrino oscillations}},}\ }\href {\doibase
  10.1038/s41586-020-2177-0} {\bibfield  {journal} {\bibinfo  {journal}
  {Nature}\ }\textbf {\bibinfo {volume} {580}},\ \bibinfo {pages} {339}
  (\bibinfo {year} {2020})},\ \bibinfo {note} {[Erratum: Nature 583, E16
  (2020)]},\ \Eprint {http://arxiv.org/abs/1910.03887}{arXiv:1910.03887
  [hep-ex]}\BibitemShut {NoStop}%
\bibitem [{\citenamefont {Hyper-Kamiokande Proto-Collaboration}\ \emph
  {et~al.}(2018)\citenamefont {Hyper-Kamiokande Proto-Collaboration},  \emph
  {et~al.}}]{protocollaboration2018hyperkamiokande}%
  \BibitemOpen
  \bibfield  {author} {\bibinfo {author} {\bibfnamefont {K.~A.}\ \bibnamefont
  {Hyper-Kamiokande Proto-Collaboration}}, ,  \emph {et~al.},\ }\bibfield
  {title} {\enquote {\bibinfo {title} {Hyper-kamiokande design report},}\
  }\href@noop {} {\  (\bibinfo {year} {2018})},\ \Eprint
  {http://arxiv.org/abs/1805.04163}{arXiv:1805.04163
  [physics.ins-det]}\BibitemShut {NoStop}%
\bibitem [{\citenamefont {Abi}\ \emph {et~al.}(2021)\citenamefont {Abi} \emph
  {et~al.}}]{DUNE:2020lwj}%
  \BibitemOpen
  \bibfield  {author} {\bibinfo {author} {\bibfnamefont {B.}~\bibnamefont
  {Abi}} \emph {et~al.} (\bibinfo {collaboration} {DUNE}),\ }\bibfield  {title}
  {\enquote {\bibinfo {title} {{Experiment Simulation Configurations
  Approximating DUNE TDR}},}\ }\href@noop {} {\  (\bibinfo {year} {2021})},\
  \Eprint {http://arxiv.org/abs/2103.04797}{arXiv:2103.04797
  [hep-ex]}\BibitemShut {NoStop}%
\bibitem [{\citenamefont {Esteban}\ \emph {et~al.}(2020)\citenamefont
  {Esteban}, \citenamefont {Gonzalez-Garcia}, \citenamefont {Maltoni},
  \citenamefont {Schwetz},\ and\ \citenamefont {Zhou}}]{Esteban:2020cvm}%
  \BibitemOpen
  \bibfield  {author} {\bibinfo {author} {\bibfnamefont {I.}~\bibnamefont
  {Esteban}}, \bibinfo {author} {\bibfnamefont {M.~C.}\ \bibnamefont
  {Gonzalez-Garcia}}, \bibinfo {author} {\bibfnamefont {M.}~\bibnamefont
  {Maltoni}}, \bibinfo {author} {\bibfnamefont {T.}~\bibnamefont {Schwetz}}, \
  and\ \bibinfo {author} {\bibfnamefont {A.}~\bibnamefont {Zhou}},\ }\bibfield
  {title} {\enquote {\bibinfo {title} {{The fate of hints: updated global
  analysis of three-flavor neutrino oscillations}},}\ }\href {\doibase
  10.1007/JHEP09(2020)178} {\bibfield  {journal} {\bibinfo  {journal} {JHEP}\
  }\textbf {\bibinfo {volume} {09}},\ \bibinfo {pages} {178} (\bibinfo {year}
  {2020})},\ \Eprint {http://arxiv.org/abs/2007.14792}{arXiv:2007.14792
  [hep-ph]}\BibitemShut {NoStop}%
\bibitem [{\citenamefont {Mohapatra}\ and\ \citenamefont
  {Smirnov}(2006)}]{Mohapatra:2006gs}%
  \BibitemOpen
  \bibfield  {author} {\bibinfo {author} {\bibfnamefont {R.~N.}\ \bibnamefont
  {Mohapatra}}\ and\ \bibinfo {author} {\bibfnamefont {A.~Y.}\ \bibnamefont
  {Smirnov}},\ }\bibfield  {title} {\enquote {\bibinfo {title} {{Neutrino Mass
  and New Physics}},}\ }\href {\doibase 10.1146/annurev.nucl.56.080805.140534}
  {\bibfield  {journal} {\bibinfo  {journal} {Ann. Rev. Nucl. Part. Sci.}\
  }\textbf {\bibinfo {volume} {56}},\ \bibinfo {pages} {569} (\bibinfo {year}
  {2006})},\ \Eprint
  {http://arxiv.org/abs/hep-ph/0603118}{arXiv:hep-ph/0603118}\BibitemShut
  {NoStop}%
\bibitem [{\citenamefont {Gehrlein}\ \emph {et~al.}(2022)\citenamefont
  {Gehrlein}, \citenamefont {Petcov}, \citenamefont {Spinrath},\ and\
  \citenamefont {Titov}}]{Gehrlein:2022nss}%
  \BibitemOpen
  \bibfield  {author} {\bibinfo {author} {\bibfnamefont {J.}~\bibnamefont
  {Gehrlein}}, \bibinfo {author} {\bibfnamefont {S.}~\bibnamefont {Petcov}},
  \bibinfo {author} {\bibfnamefont {M.}~\bibnamefont {Spinrath}}, \ and\
  \bibinfo {author} {\bibfnamefont {A.}~\bibnamefont {Titov}},\ }\bibfield
  {title} {\enquote {\bibinfo {title} {{Testing neutrino flavor models}},}\
  }in\ \href@noop {} {\emph {\bibinfo {booktitle} {{Snowmass 2021}}}}\
  (\bibinfo {year} {2022})\ \Eprint
  {http://arxiv.org/abs/2203.06219}{arXiv:2203.06219 [hep-ph]}\BibitemShut
  {NoStop}%
\bibitem [{\citenamefont {Adamson}\ \emph {et~al.}(2020)\citenamefont {Adamson}
  \emph {et~al.}}]{MINOS:2020llm}%
  \BibitemOpen
  \bibfield  {author} {\bibinfo {author} {\bibfnamefont {P.}~\bibnamefont
  {Adamson}} \emph {et~al.} (\bibinfo {collaboration} {MINOS+}),\ }\bibfield
  {title} {\enquote {\bibinfo {title} {{Precision Constraints for Three-Flavor
  Neutrino Oscillations from the Full MINOS+ and MINOS Dataset}},}\ }\href
  {\doibase 10.1103/PhysRevLett.125.131802} {\bibfield  {journal} {\bibinfo
  {journal} {Phys. Rev. Lett.}\ }\textbf {\bibinfo {volume} {125}},\ \bibinfo
  {pages} {131802} (\bibinfo {year} {2020})},\ \Eprint
  {http://arxiv.org/abs/2006.15208}{arXiv:2006.15208 [hep-ex]}\BibitemShut
  {NoStop}%
\bibitem [{\citenamefont {Wendell}\ \emph {et~al.}(2010)\citenamefont {Wendell}
  \emph {et~al.}}]{Super-Kamiokande:2010orq}%
  \BibitemOpen
  \bibfield  {author} {\bibinfo {author} {\bibfnamefont {R.}~\bibnamefont
  {Wendell}} \emph {et~al.} (\bibinfo {collaboration} {Super-Kamiokande}),\
  }\bibfield  {title} {\enquote {\bibinfo {title} {{Atmospheric neutrino
  oscillation analysis with sub-leading effects in Super-Kamiokande I, II, and
  III}},}\ }\href {\doibase 10.1103/PhysRevD.81.092004} {\bibfield  {journal}
  {\bibinfo  {journal} {Phys. Rev. D}\ }\textbf {\bibinfo {volume} {81}},\
  \bibinfo {pages} {092004} (\bibinfo {year} {2010})},\ \Eprint
  {http://arxiv.org/abs/1002.3471}{arXiv:1002.3471 [hep-ex]}\BibitemShut
  {NoStop}%
\bibitem [{\citenamefont {Acero}\ \emph
  {et~al.}(2022{\natexlab{b}})\citenamefont {Acero} \emph
  {et~al.}}]{NOvA:2021nfi}%
  \BibitemOpen
  \bibfield  {author} {\bibinfo {author} {\bibfnamefont {M.~A.}\ \bibnamefont
  {Acero}} \emph {et~al.} (\bibinfo {collaboration} {NO$\nu$A}),\ }\bibfield
  {title} {\enquote {\bibinfo {title} {{Improved measurement of neutrino
  oscillation parameters by the NO$\nu$A experiment}},}\ }\href {\doibase
  10.1103/PhysRevD.106.032004} {\bibfield  {journal} {\bibinfo  {journal}
  {Phys. Rev. D}\ }\textbf {\bibinfo {volume} {106}},\ \bibinfo {pages}
  {032004} (\bibinfo {year} {2022}{\natexlab{b}})},\ \Eprint
  {http://arxiv.org/abs/2108.08219}{arXiv:2108.08219 [hep-ex]}\BibitemShut
  {NoStop}%
\bibitem [{\citenamefont {Jiang}\ \emph {et~al.}(2019)\citenamefont {Jiang}
  \emph {et~al.}}]{Super-Kamiokande:2019gzr}%
  \BibitemOpen
  \bibfield  {author} {\bibinfo {author} {\bibfnamefont {M.}~\bibnamefont
  {Jiang}} \emph {et~al.} (\bibinfo {collaboration} {Super-Kamiokande}),\
  }\bibfield  {title} {\enquote {\bibinfo {title} {{Atmospheric Neutrino
  Oscillation Analysis with Improved Event Reconstruction in Super-Kamiokande
  IV}},}\ }\href {\doibase 10.1093/ptep/ptz015} {\bibfield  {journal} {\bibinfo
   {journal} {PTEP}\ }\textbf {\bibinfo {volume} {2019}},\ \bibinfo {pages}
  {053F01} (\bibinfo {year} {2019})},\ \Eprint
  {http://arxiv.org/abs/1901.03230}{arXiv:1901.03230 [hep-ex]}\BibitemShut
  {NoStop}%
\bibitem [{\citenamefont {Abbasi}\ \emph {et~al.}(2023)\citenamefont {Abbasi}
  \emph {et~al.}}]{IceCubeCollaboration:2023wtb}%
  \BibitemOpen
  \bibfield  {author} {\bibinfo {author} {\bibfnamefont {R.}~\bibnamefont
  {Abbasi}} \emph {et~al.} (\bibinfo {collaboration} {(IceCube Collaboration)*,
  IceCube}),\ }\bibfield  {title} {\enquote {\bibinfo {title} {{Measurement of
  atmospheric neutrino mixing with improved IceCube DeepCore calibration and
  data processing}},}\ }\href {\doibase 10.1103/PhysRevD.108.012014} {\bibfield
   {journal} {\bibinfo  {journal} {Phys. Rev. D}\ }\textbf {\bibinfo {volume}
  {108}},\ \bibinfo {pages} {012014} (\bibinfo {year} {2023})},\ \Eprint
  {http://arxiv.org/abs/2304.12236}{arXiv:2304.12236 [hep-ex]}\BibitemShut
  {NoStop}%
\bibitem [{\citenamefont {Abe}\ \emph {et~al.}(2011)\citenamefont {Abe} \emph
  {et~al.}}]{Abe:2011ks}%
  \BibitemOpen
  \bibfield  {author} {\bibinfo {author} {\bibfnamefont {K.}~\bibnamefont
  {Abe}} \emph {et~al.} (\bibinfo {collaboration} {T2K Collaboration}),\
  }\bibfield  {title} {\enquote {\bibinfo {title} {The {T2K} experiment},}\
  }\href {\doibase 10.1016/j.nima.2011.06.067} {\bibfield  {journal} {\bibinfo
  {journal} {Nucl. Instrum. Methods Phys. Res. Sect. A}\ }\textbf {\bibinfo
  {volume} {659}},\ \bibinfo {pages} {106} (\bibinfo {year} {2011})},\ \Eprint
  {http://arxiv.org/abs/1106.1238}{arXiv:1106.1238
  [physics.ins-det]}\BibitemShut {NoStop}%
\bibitem [{\citenamefont {Ayres}\ \emph {et~al.}(2007)\citenamefont {Ayres}
  \emph {et~al.}}]{ayres2007nova}%
  \BibitemOpen
  \bibfield  {author} {\bibinfo {author} {\bibfnamefont {D.}~\bibnamefont
  {Ayres}} \emph {et~al.} (\bibinfo {collaboration} {NO$\nu$A Collaboration}),\
  }\bibfield  {title} {\enquote {\bibinfo {title} {{The NO$\nu$A Technical
  Design Report}},}\ }\href {\doibase 10.2172/935497} {\  (\bibinfo {year}
  {2007}),\ 10.2172/935497}\BibitemShut {NoStop}%
\bibitem [{\citenamefont {Abe}\ \emph {et~al.}(2014)\citenamefont {Abe} \emph
  {et~al.}}]{T2K:2013ppw}%
  \BibitemOpen
  \bibfield  {author} {\bibinfo {author} {\bibfnamefont {K.}~\bibnamefont
  {Abe}} \emph {et~al.} (\bibinfo {collaboration} {T2K}),\ }\bibfield  {title}
  {\enquote {\bibinfo {title} {{Observation of Electron Neutrino Appearance in
  a Muon Neutrino Beam}},}\ }\href {\doibase 10.1103/PhysRevLett.112.061802}
  {\bibfield  {journal} {\bibinfo  {journal} {Phys. Rev. Lett.}\ }\textbf
  {\bibinfo {volume} {112}},\ \bibinfo {pages} {061802} (\bibinfo {year}
  {2014})},\ \Eprint {http://arxiv.org/abs/1311.4750}{arXiv:1311.4750
  [hep-ex]}\BibitemShut {NoStop}%
\bibitem [{\citenamefont {Zhang}\ and\ \citenamefont
  {Cao}(2023)}]{Zhang:2022zoc}%
  \BibitemOpen
  \bibfield  {author} {\bibinfo {author} {\bibfnamefont {J.}~\bibnamefont
  {Zhang}}\ and\ \bibinfo {author} {\bibfnamefont {J.}~\bibnamefont {Cao}},\
  }\bibfield  {title} {\enquote {\bibinfo {title} {{Towards a sub-percent
  precision measurement of sin$^{2}$\ensuremath{\theta}$_{13}$ with reactor
  antineutrinos}},}\ }\href {\doibase 10.1007/JHEP03(2023)072} {\bibfield
  {journal} {\bibinfo  {journal} {JHEP}\ }\textbf {\bibinfo {volume} {03}},\
  \bibinfo {pages} {072} (\bibinfo {year} {2023})},\ \Eprint
  {http://arxiv.org/abs/2206.15317}{arXiv:2206.15317 [hep-ex]}\BibitemShut
  {NoStop}%
\bibitem [{\citenamefont {Strait}\ \emph {et~al.}(2016)\citenamefont {Strait}
  \emph {et~al.}}]{DUNE:2016evb}%
  \BibitemOpen
  \bibfield  {author} {\bibinfo {author} {\bibfnamefont {J.}~\bibnamefont
  {Strait}} \emph {et~al.} (\bibinfo {collaboration} {DUNE}),\ }\bibfield
  {title} {\enquote {\bibinfo {title} {{Long-Baseline Neutrino Facility (LBNF)
  and Deep Underground Neutrino Experiment (DUNE)}: {Conceptual Design Report,
  Volume 3: Long-Baseline Neutrino Facility for DUNE June 24, 2015}},}\
  }\href@noop {} {\  (\bibinfo {year} {2016})},\ \Eprint
  {http://arxiv.org/abs/1601.05823}{arXiv:1601.05823
  [physics.ins-det]}\BibitemShut {NoStop}%
\bibitem [{\citenamefont {Fukasawa}\ \emph {et~al.}(2017)\citenamefont
  {Fukasawa}, \citenamefont {Ghosh},\ and\ \citenamefont
  {Yasuda}}]{Fukasawa:2016yue}%
  \BibitemOpen
  \bibfield  {author} {\bibinfo {author} {\bibfnamefont {S.}~\bibnamefont
  {Fukasawa}}, \bibinfo {author} {\bibfnamefont {M.}~\bibnamefont {Ghosh}}, \
  and\ \bibinfo {author} {\bibfnamefont {O.}~\bibnamefont {Yasuda}},\
  }\bibfield  {title} {\enquote {\bibinfo {title} {{Complementarity Between
  Hyperkamiokande and DUNE in Determining Neutrino Oscillation Parameters}},}\
  }\href {\doibase 10.1016/j.nuclphysb.2017.02.008} {\bibfield  {journal}
  {\bibinfo  {journal} {Nucl. Phys. B}\ }\textbf {\bibinfo {volume} {918}},\
  \bibinfo {pages} {337} (\bibinfo {year} {2017})},\ \Eprint
  {http://arxiv.org/abs/1607.03758}{arXiv:1607.03758 [hep-ph]}\BibitemShut
  {NoStop}%
\bibitem [{\citenamefont {Agarwalla}\ \emph {et~al.}(2018)\citenamefont
  {Agarwalla}, \citenamefont {Chatterjee}, \citenamefont {Petcov},\ and\
  \citenamefont {Titov}}]{Agarwalla:2017wct}%
  \BibitemOpen
  \bibfield  {author} {\bibinfo {author} {\bibfnamefont {S.~K.}\ \bibnamefont
  {Agarwalla}}, \bibinfo {author} {\bibfnamefont {S.~S.}\ \bibnamefont
  {Chatterjee}}, \bibinfo {author} {\bibfnamefont {S.~T.}\ \bibnamefont
  {Petcov}}, \ and\ \bibinfo {author} {\bibfnamefont {A.~V.}\ \bibnamefont
  {Titov}},\ }\bibfield  {title} {\enquote {\bibinfo {title} {{Addressing
  Neutrino Mixing Models with DUNE and T2HK}},}\ }\href {\doibase
  10.1140/epjc/s10052-018-5772-6} {\bibfield  {journal} {\bibinfo  {journal}
  {Eur. Phys. J. C}\ }\textbf {\bibinfo {volume} {78}},\ \bibinfo {pages} {286}
  (\bibinfo {year} {2018})},\ \Eprint
  {http://arxiv.org/abs/1711.02107}{arXiv:1711.02107 [hep-ph]}\BibitemShut
  {NoStop}%
\bibitem [{\citenamefont {Rosauro~Alcaraz}\ \emph {et~al.}(2022)\citenamefont
  {Rosauro~Alcaraz}, \citenamefont {Blennow}, \citenamefont
  {Fernandez-Martinez},\ and\ \citenamefont {Ota}}]{RosauroAlcaraz:2022str}%
  \BibitemOpen
  \bibfield  {author} {\bibinfo {author} {\bibfnamefont {S.}~\bibnamefont
  {Rosauro~Alcaraz}}, \bibinfo {author} {\bibfnamefont {M.}~\bibnamefont
  {Blennow}}, \bibinfo {author} {\bibfnamefont {E.}~\bibnamefont
  {Fernandez-Martinez}}, \ and\ \bibinfo {author} {\bibfnamefont
  {T.}~\bibnamefont {Ota}},\ }\bibfield  {title} {\enquote {\bibinfo {title}
  {{Physics potential of the ESSnuSB}},}\ }\href {\doibase 10.22323/1.402.0063}
  {\bibfield  {journal} {\bibinfo  {journal} {PoS}\ }\textbf {\bibinfo {volume}
  {NuFact2021}},\ \bibinfo {pages} {063} (\bibinfo {year} {2022})}\BibitemShut
  {NoStop}%
\bibitem [{\citenamefont {Blennow}\ \emph {et~al.}(2020)\citenamefont
  {Blennow}, \citenamefont {Ghosh}, \citenamefont {Ohlsson},\ and\
  \citenamefont {Titov}}]{Blennow:2020snb}%
  \BibitemOpen
  \bibfield  {author} {\bibinfo {author} {\bibfnamefont {M.}~\bibnamefont
  {Blennow}}, \bibinfo {author} {\bibfnamefont {M.}~\bibnamefont {Ghosh}},
  \bibinfo {author} {\bibfnamefont {T.}~\bibnamefont {Ohlsson}}, \ and\
  \bibinfo {author} {\bibfnamefont {A.}~\bibnamefont {Titov}},\ }\bibfield
  {title} {\enquote {\bibinfo {title} {{Testing Lepton Flavor Models at
  ESSnuSB}},}\ }\href {\doibase 10.1007/JHEP07(2020)014} {\bibfield  {journal}
  {\bibinfo  {journal} {JHEP}\ }\textbf {\bibinfo {volume} {07}},\ \bibinfo
  {pages} {014} (\bibinfo {year} {2020})},\ \Eprint
  {http://arxiv.org/abs/2004.00017}{arXiv:2004.00017 [hep-ph]}\BibitemShut
  {NoStop}%
\bibitem [{\citenamefont {Geer}(1998)}]{Geer:1997iz}%
  \BibitemOpen
  \bibfield  {author} {\bibinfo {author} {\bibfnamefont {S.}~\bibnamefont
  {Geer}},\ }\bibfield  {title} {\enquote {\bibinfo {title} {{Neutrino beams
  from muon storage rings: Characteristics and physics potential}},}\ }\href
  {\doibase 10.1103/PhysRevD.57.6989} {\bibfield  {journal} {\bibinfo
  {journal} {Phys. Rev. D}\ }\textbf {\bibinfo {volume} {57}},\ \bibinfo
  {pages} {6989} (\bibinfo {year} {1998})},\ \bibinfo {note} {[Erratum:
  Phys.Rev.D 59, 039903 (1999)]},\ \Eprint
  {http://arxiv.org/abs/hep-ph/9712290}{arXiv:hep-ph/9712290}\BibitemShut
  {NoStop}%
\bibitem [{\citenamefont {Choubey}\ \emph {et~al.}(2011)\citenamefont {Choubey}
  \emph {et~al.}}]{IDS-NF:2011swj}%
  \BibitemOpen
  \bibfield  {author} {\bibinfo {author} {\bibfnamefont {S.}~\bibnamefont
  {Choubey}} \emph {et~al.} (\bibinfo {collaboration} {IDS-NF}),\ }\bibfield
  {title} {\enquote {\bibinfo {title} {{International Design Study for the
  Neutrino Factory, Interim Design Report}},}\ }\href@noop {} {\  (\bibinfo
  {year} {2011})},\ \Eprint {http://arxiv.org/abs/1112.2853}{arXiv:1112.2853
  [hep-ex]}\BibitemShut {NoStop}%
\bibitem [{\citenamefont {Bogacz}\ \emph {et~al.}(2022)\citenamefont {Bogacz}
  \emph {et~al.}}]{Bogacz:2022xsj}%
  \BibitemOpen
  \bibfield  {author} {\bibinfo {author} {\bibfnamefont {A.}~\bibnamefont
  {Bogacz}} \emph {et~al.},\ }\bibfield  {title} {\enquote {\bibinfo {title}
  {{The Physics Case for a Neutrino Factory}},}\ }in\ \href@noop {} {\emph
  {\bibinfo {booktitle} {{Snowmass 2021}}}}\ (\bibinfo {year} {2022})\ \Eprint
  {http://arxiv.org/abs/2203.08094}{arXiv:2203.08094 [hep-ph]}\BibitemShut
  {NoStop}%
\bibitem [{\citenamefont {Denton}\ and\ \citenamefont
  {Gehrlein}(2024)}]{Denton:2024glz}%
  \BibitemOpen
  \bibfield  {author} {\bibinfo {author} {\bibfnamefont {P.~B.}\ \bibnamefont
  {Denton}}\ and\ \bibinfo {author} {\bibfnamefont {J.}~\bibnamefont
  {Gehrlein}},\ }\bibfield  {title} {\enquote {\bibinfo {title} {{A Modern Look
  at the Oscillation Physics Case for a Neutrino Factory}},}\ }\href@noop {} {\
   (\bibinfo {year} {2024})},\ \Eprint
  {http://arxiv.org/abs/2407.02572}{arXiv:2407.02572 [hep-ph]}\BibitemShut
  {NoStop}%
\bibitem [{\citenamefont {Huber}\ \emph {et~al.}(2005)\citenamefont {Huber},
  \citenamefont {Lindner},\ and\ \citenamefont {Winter}}]{Huber:2004ka}%
  \BibitemOpen
  \bibfield  {author} {\bibinfo {author} {\bibfnamefont {P.}~\bibnamefont
  {Huber}}, \bibinfo {author} {\bibfnamefont {M.}~\bibnamefont {Lindner}}, \
  and\ \bibinfo {author} {\bibfnamefont {W.}~\bibnamefont {Winter}},\
  }\bibfield  {title} {\enquote {\bibinfo {title} {{Simulation of long-baseline
  neutrino oscillation experiments with GLoBES (General Long Baseline
  Experiment Simulator)}},}\ }\href {\doibase 10.1016/j.cpc.2005.01.003}
  {\bibfield  {journal} {\bibinfo  {journal} {Comput. Phys. Commun.}\ }\textbf
  {\bibinfo {volume} {167}},\ \bibinfo {pages} {195} (\bibinfo {year}
  {2005})},\ \Eprint
  {http://arxiv.org/abs/hep-ph/0407333}{arXiv:hep-ph/0407333}\BibitemShut
  {NoStop}%
\bibitem [{\citenamefont {Huber}\ \emph {et~al.}(2007)\citenamefont {Huber},
  \citenamefont {Kopp}, \citenamefont {Lindner}, \citenamefont {Rolinec},\ and\
  \citenamefont {Winter}}]{Huber:2007ji}%
  \BibitemOpen
  \bibfield  {author} {\bibinfo {author} {\bibfnamefont {P.}~\bibnamefont
  {Huber}}, \bibinfo {author} {\bibfnamefont {J.}~\bibnamefont {Kopp}},
  \bibinfo {author} {\bibfnamefont {M.}~\bibnamefont {Lindner}}, \bibinfo
  {author} {\bibfnamefont {M.}~\bibnamefont {Rolinec}}, \ and\ \bibinfo
  {author} {\bibfnamefont {W.}~\bibnamefont {Winter}},\ }\bibfield  {title}
  {\enquote {\bibinfo {title} {{New features in the simulation of neutrino
  oscillation experiments with GLoBES 3.0: General Long Baseline Experiment
  Simulator}},}\ }\href {\doibase 10.1016/j.cpc.2007.05.004} {\bibfield
  {journal} {\bibinfo  {journal} {Comput. Phys. Commun.}\ }\textbf {\bibinfo
  {volume} {177}},\ \bibinfo {pages} {432} (\bibinfo {year} {2007})},\ \Eprint
  {http://arxiv.org/abs/hep-ph/0701187}{arXiv:hep-ph/0701187}\BibitemShut
  {NoStop}%
\bibitem [{\citenamefont {Abusleme}\ \emph {et~al.}(2022)\citenamefont
  {Abusleme} \emph {et~al.}}]{JUNO:2022mxj}%
  \BibitemOpen
  \bibfield  {author} {\bibinfo {author} {\bibfnamefont {A.}~\bibnamefont
  {Abusleme}} \emph {et~al.} (\bibinfo {collaboration} {JUNO}),\ }\bibfield
  {title} {\enquote {\bibinfo {title} {{Sub-percent precision measurement of
  neutrino oscillation parameters with JUNO}},}\ }\href {\doibase
  10.1088/1674-1137/ac8bc9} {\bibfield  {journal} {\bibinfo  {journal} {Chin.
  Phys. C}\ }\textbf {\bibinfo {volume} {46}},\ \bibinfo {pages} {123001}
  (\bibinfo {year} {2022})},\ \Eprint
  {http://arxiv.org/abs/2204.13249}{arXiv:2204.13249 [hep-ex]}\BibitemShut
  {NoStop}%
\bibitem [{\citenamefont {Altarelli}\ and\ \citenamefont
  {Feruglio}(2010)}]{Altarelli:2010gt}%
  \BibitemOpen
  \bibfield  {author} {\bibinfo {author} {\bibfnamefont {G.}~\bibnamefont
  {Altarelli}}\ and\ \bibinfo {author} {\bibfnamefont {F.}~\bibnamefont
  {Feruglio}},\ }\bibfield  {title} {\enquote {\bibinfo {title} {{Discrete
  Flavor Symmetries and Models of Neutrino Mixing}},}\ }\href {\doibase
  10.1103/RevModPhys.82.2701} {\bibfield  {journal} {\bibinfo  {journal} {Rev.
  Mod. Phys.}\ }\textbf {\bibinfo {volume} {82}},\ \bibinfo {pages} {2701}
  (\bibinfo {year} {2010})},\ \Eprint
  {http://arxiv.org/abs/1002.0211}{arXiv:1002.0211 [hep-ph]}\BibitemShut
  {NoStop}%
\bibitem [{\citenamefont {King}\ and\ \citenamefont
  {Luhn}(2013)}]{King:2013eh}%
  \BibitemOpen
  \bibfield  {author} {\bibinfo {author} {\bibfnamefont {S.~F.}\ \bibnamefont
  {King}}\ and\ \bibinfo {author} {\bibfnamefont {C.}~\bibnamefont {Luhn}},\
  }\bibfield  {title} {\enquote {\bibinfo {title} {{Neutrino Mass and Mixing
  with Discrete Symmetry}},}\ }\href {\doibase 10.1088/0034-4885/76/5/056201}
  {\bibfield  {journal} {\bibinfo  {journal} {Rept. Prog. Phys.}\ }\textbf
  {\bibinfo {volume} {76}},\ \bibinfo {pages} {056201} (\bibinfo {year}
  {2013})},\ \Eprint {http://arxiv.org/abs/1301.1340}{arXiv:1301.1340
  [hep-ph]}\BibitemShut {NoStop}%
\bibitem [{\citenamefont {Petcov}(2015)}]{Petcov:2014laa}%
  \BibitemOpen
  \bibfield  {author} {\bibinfo {author} {\bibfnamefont {S.~T.}\ \bibnamefont
  {Petcov}},\ }\bibfield  {title} {\enquote {\bibinfo {title} {{Predicting the
  values of the leptonic CP violation phases in theories with discrete flavour
  symmetries}},}\ }\href {\doibase 10.1016/j.nuclphysb.2015.01.011} {\bibfield
  {journal} {\bibinfo  {journal} {Nucl. Phys. B}\ }\textbf {\bibinfo {volume}
  {892}},\ \bibinfo {pages} {400} (\bibinfo {year} {2015})},\ \Eprint
  {http://arxiv.org/abs/1405.6006}{arXiv:1405.6006 [hep-ph]}\BibitemShut
  {NoStop}%
\bibitem [{\citenamefont {Costa}\ and\ \citenamefont
  {King}(2023)}]{Costa:2023bxw}%
  \BibitemOpen
  \bibfield  {author} {\bibinfo {author} {\bibfnamefont {F.}~\bibnamefont
  {Costa}}\ and\ \bibinfo {author} {\bibfnamefont {S.~F.}\ \bibnamefont
  {King}},\ }\bibfield  {title} {\enquote {\bibinfo {title} {{Neutrino Mixing
  Sum Rules and the Littlest Seesaw}},}\ }\href {\doibase
  10.3390/universe9110472} {\bibfield  {journal} {\bibinfo  {journal}
  {Universe}\ }\textbf {\bibinfo {volume} {9}},\ \bibinfo {pages} {472}
  (\bibinfo {year} {2023})},\ \Eprint
  {http://arxiv.org/abs/2307.13895}{arXiv:2307.13895 [hep-ph]}\BibitemShut
  {NoStop}%
\end{thebibliography}%

\appendix
\section{Simulated event rates of T2HK}\label{app:t2hk}
\begin{figure*}
\includegraphics[width=0.45\textwidth]{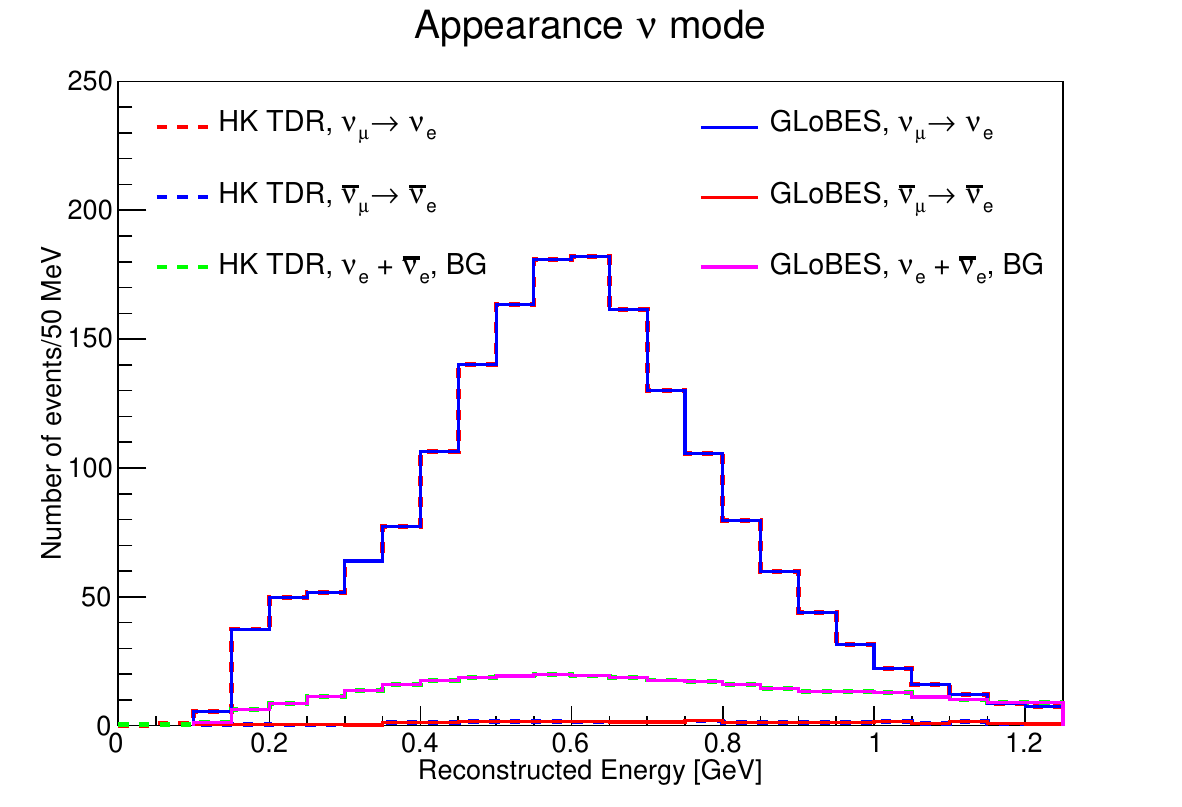}
\includegraphics[width=0.45\textwidth]{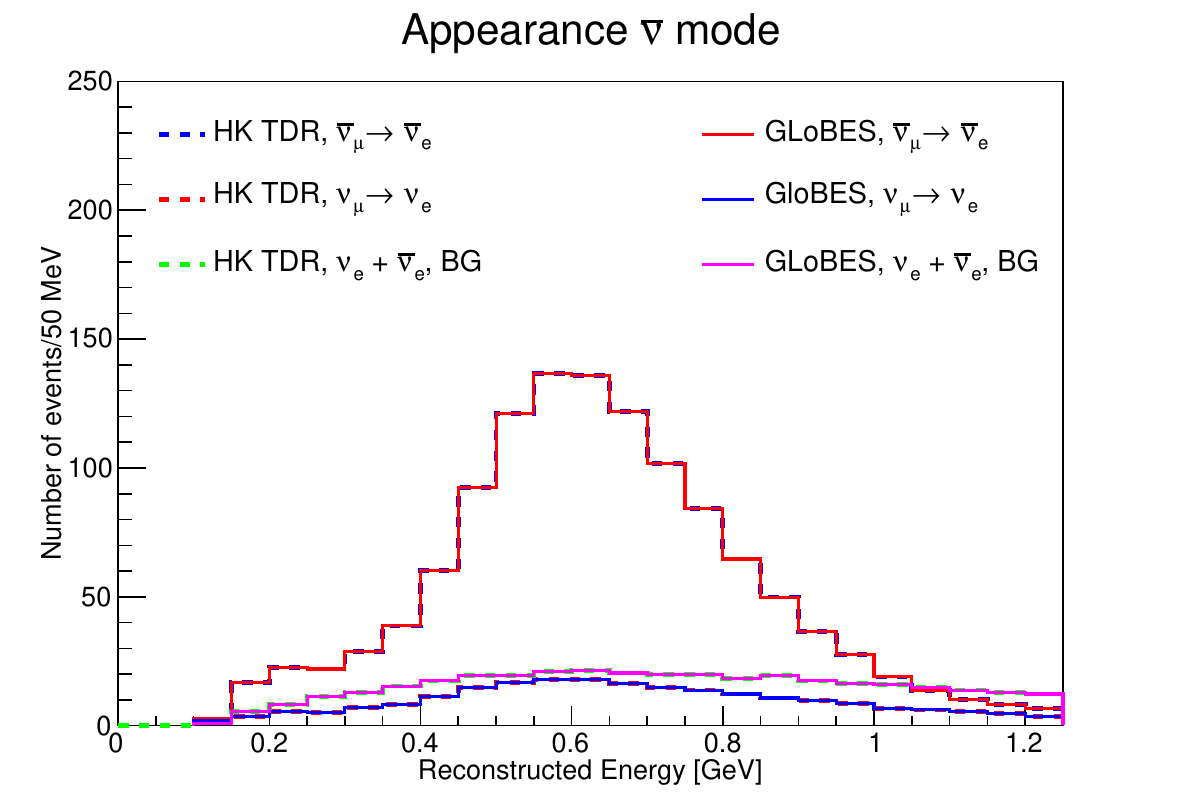}
\includegraphics[width=0.45\textwidth]{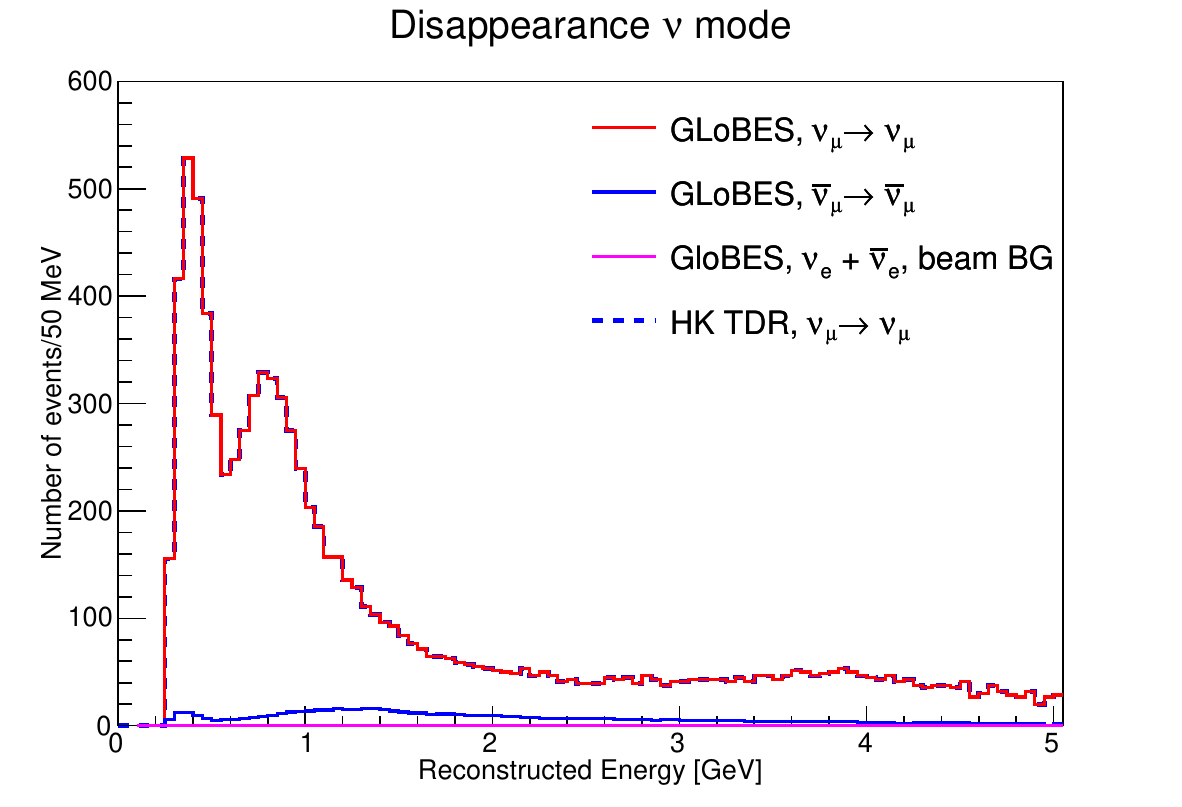}
\includegraphics[width=0.45\textwidth]{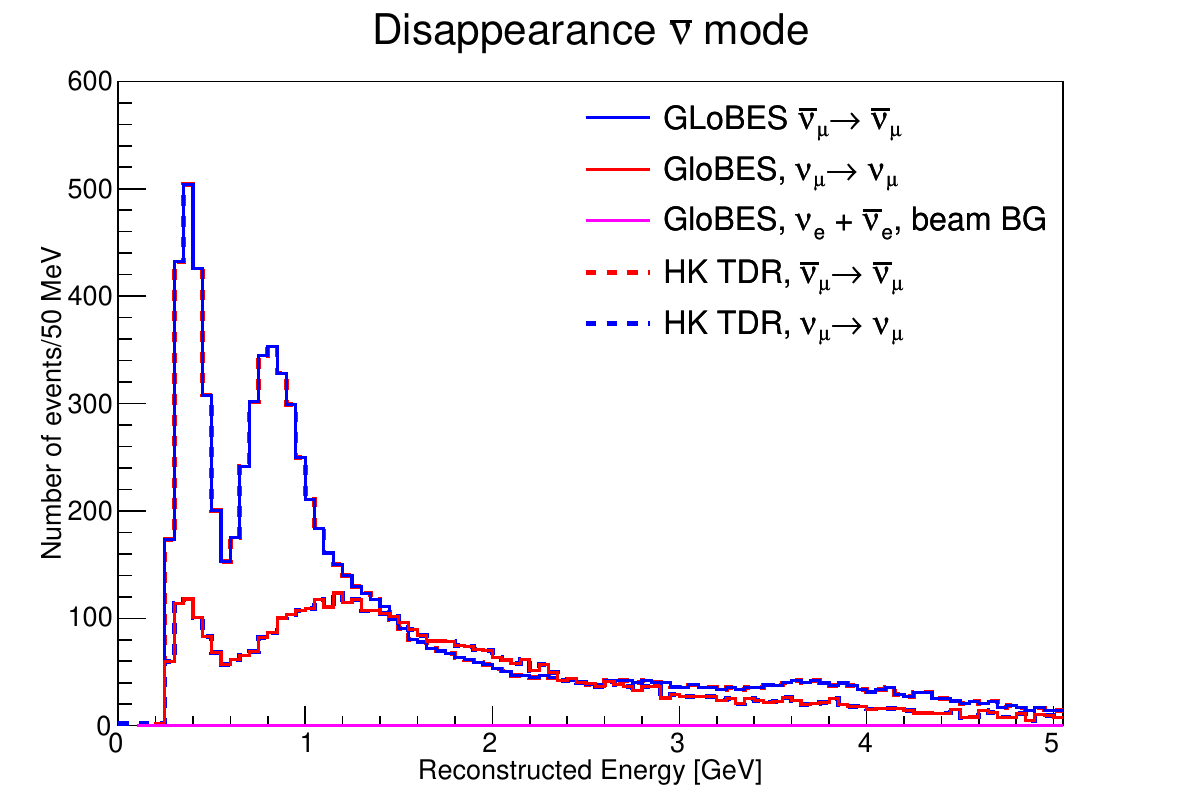}
\caption{\label{fig:hkeventrate} The event spectra for the signal and background in the T2HK simulated sample are presented as a function of reconstructed neutrino energy. The upper (lower) spectrum corresponds to the \emph{appearance} (\emph{disappearance}) channel, while the left (right) spectrum involves to the $\nu$-mode ($\overline{\nu}$-mode) respectively. \emph{Normal} mass ordering (MO), $\delta_{CP} = 0$, and other oscillation parameters provided in the Technical Design Report (TDR) ~\cite{protocollaboration2018hyperkamiokande} are assumed for the rate calculation.}
\end{figure*}

The simulated experimental configuration of T2HK, as outlined in the technical design report (TDR)~\cite{protocollaboration2018hyperkamiokande}, is presented in Table~\ref{tab:hkconfig}, along with basic features listed in Table~\ref{tab:t2hkdune}. The event rate of four simulated data samples for neutrino oscillation measurements, closely aligned with the T2HK TDR, is presented in Fig.~\ref{fig:hkeventrate}. This study employs a tuning of detection efficiency as a function of energy to match with the TDR report, enabling us to closely approach the physical potential of T2HK in measuring neutrino oscillation parameters. 
\begin{table*}
    \caption{\label{tab:hkconfig}T2HK specifications for simulation, based on Ref~\cite{protocollaboration2018hyperkamiokande}, along with Table~\ref{tab:t2hkdune} }
   \begin{ruledtabular}
   \begin{tabular}{lll}    \textbf{Characteristics} & \textbf{T2HK experimental setup for simulation} ~\cite{protocollaboration2018hyperkamiokande}  \\\hline
   Matter density & 2.6 $gcc^{-1}$ \\
  
  Signal and background efficiency \\
   \emph{--Appearance} \\
    \textit{$\nu$ mode} ($\nu_{\mu}\rightarrow\nu_e$) & 61.6\% $\nu_e$ QE, 48.5\% $\overline{\nu}_e$ QE\\
     & 0.076\% $\nu_\mu$ CC, 23.6\% $\nu_e$ beam and 12.1\% $\overline{\nu}_e$ beam, 0.31\% $\nu_\mu$ NC \\
    \textit{$\overline{\nu}$ mode ($\overline{\nu}_\mu\rightarrow\overline{\nu}_e$)} & 68.6\% $\overline{\nu}_e$ QE, 43.6\% $\nu_e$ QE\\
       & 0.023\% $\nu_{\mu}$ CC, 0.015\% $\overline{\nu}_{\mu}$ CC, 13\% $\nu_e$ beam and 29.6$\overline{\nu}_e$ beam, 0.9\% $\overline{\nu}_\mu$ NC\\ 
   \emph{--Disappearance}\\
    \textit{$\nu$ mode ($\nu_{\mu}\rightarrow\nu_\mu$)} & (69.7\%,15.7\%)\footnote{(CCQE,CC-nonQE) where CC : Charged-Current and QE : Quasi-Elastic scattering} $\nu_\mu$, (73.2\%, 40.6\%)$\overline{\nu}_\mu$ \\
    &1.01\%$\nu_{e}$ CC, 0.42\%$\nu_e$ beam and 0.6\%$\overline{\nu}_e$ beam, 1.1\% $\nu_{\mu}$NC
    \textit{$\overline{\nu}$ mode  ($\overline{\nu}_\mu\rightarrow\overline{\nu}_\mu$)} \\
    \textit{$\overline{\nu}$ mode ($\overline{\nu}_\mu\rightarrow\overline{\nu}_\mu$)} & (73.1\%,41.5\%) $\overline{\nu}_\mu$, (67.4\%,15.2\%) $\nu_\mu$\\
   &0.46\% $\nu_{e}$CC, 0.09\% $\nu_e$ beam and 0.3\%$\overline{\nu}_e$ beam, 2.79\% $\overline{\nu}_\mu$NC \\
    Systematics & \\
    \emph{--Signal}\footnote{normalization (calibration) error for signal} & 3.2\% (2.5\%) $\nu_e$ appearance, 3.9\% (2.5\%) $\overline{\nu}_e$ appearance\\
    &3.6\% (2.5\%) $\nu_{\mu}$ disappearance, 3.6\% (2.5\%) $\overline{\nu}_\mu$ disappearance \\
    \emph{--Background}\footnote{normalization (calibration) error for background} & 10\% (2.5\%)\\
    Energy resolution & $0.03/\sqrt{E (GeV)}$\\
   Energy window & 0.10 : 1.30 GeV (\textit{APP}), 0.20 : 5.05 GeV (\textit{DIS})\\
  \end{tabular}
    \end{ruledtabular}
\end{table*}

\section{ Optimization of ESSnuSB and Neutrino Factory}\label{app:essnusbnf}
\begin{figure*}
\includegraphics[width=0.45\textwidth]
{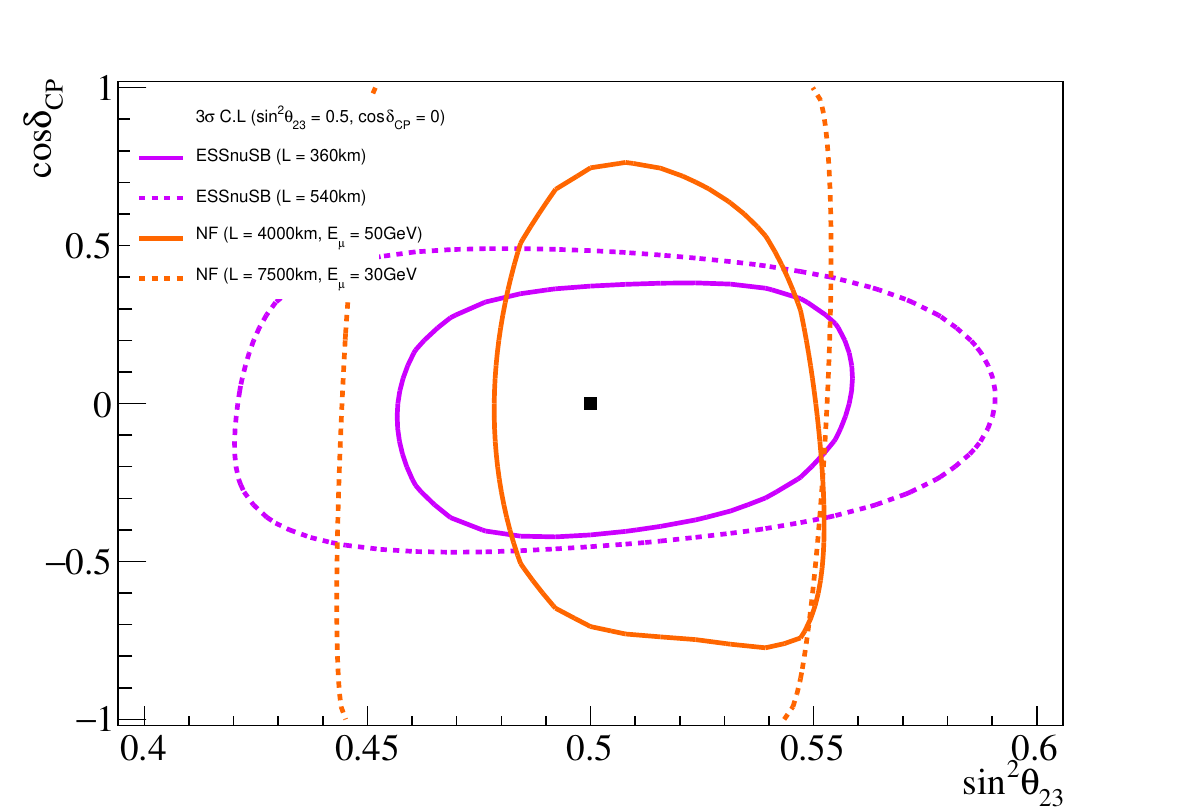}
\includegraphics[width=0.45\textwidth]
{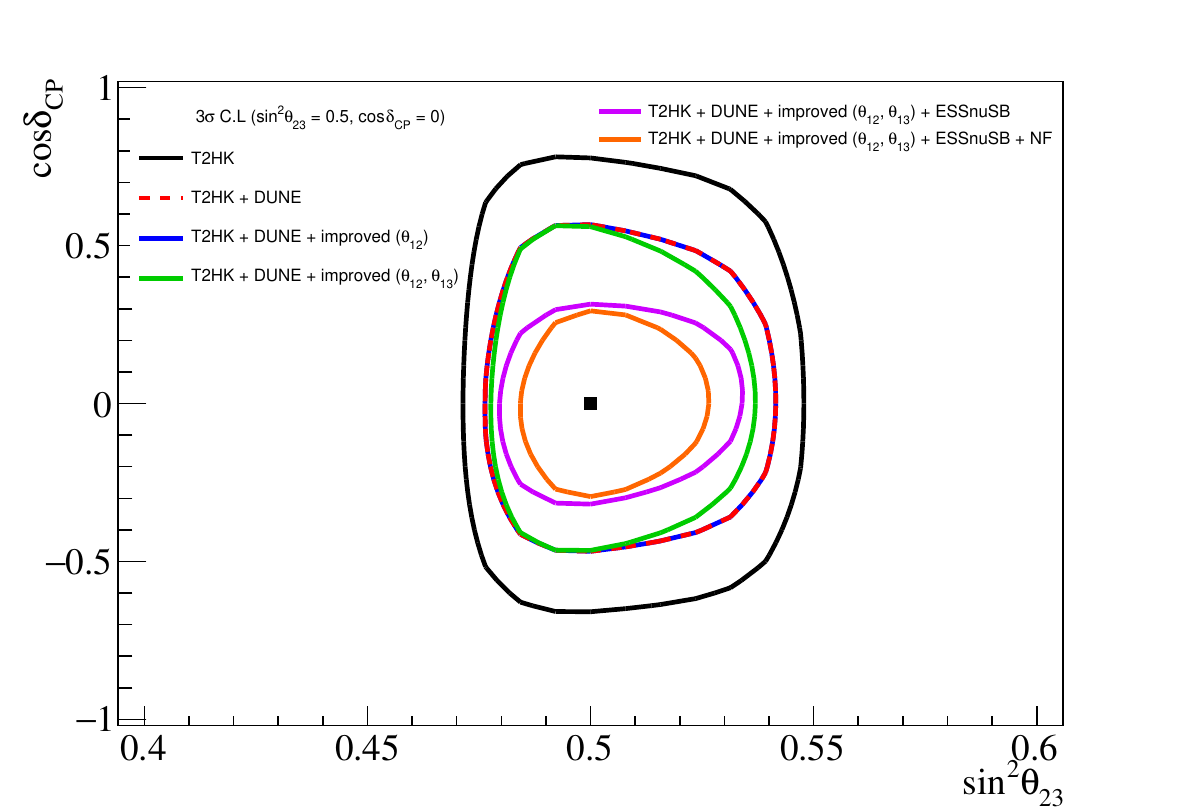}
\caption{\label{fig:cosdeltacp_sensi_exps}Left shows comparision of allowed parameter space in ($\cos\delta_{CP}, \sin^{2}\theta_{23}$) using ESSnuSB and NF with different baseline options. The sensitivity of joint analyses involving T2HK and DUNE to $\delta_{CP} - \theta_{23}$  measurements is illustrated on the right, utilizing ESSnuSB with baseline of 360km and NF with 4000~km baseline and 50~GeV of stored muons. Study is done at $\cos\delta_{CP}=0$ and $\sin^{2}\theta_{23}=0.5$, while the other parameters are constrained by the global fit data.}
\end{figure*}

Both ESSnuSB and Neutrino Factory exhibit some flexibility in their current designs. Our selection is grounded in the optimal sensitivity achieved concurrently on both \dcp, specifically the precision of $\cos \delta_{CP}$ and $\sin^2\theta_{23}$, which are crucial for evaluating the class of flavor models discussed in Section~.\ref{sec:flavormodel}. We examine the physical sensitivity of these parameters in the context of ESSnuSB, utilizing two different detector baseline options. The left plot of Fig.~\ref{fig:cosdeltacp_sensi_exps} demonstrates that a baseline of 360 km yields superior performance compared to a baseline of 540 km. This study examines two scenarios for the Neutrino Factory: one with a muon energy of 50 GeV and a baseline of 4000 km, and another with a muon energy of 30 GeV and the same baseline of 4000 km. The left plot of Fig.~\ref{fig:cosdeltacp_sensi_exps} clearly indicates that the former scenario is anticipated to provide improved sensitivity to both $\cos\delta_{CP}$ and $\sin^2\theta_{23}$. Furthermore, it has been observed that ESSnuSB significantly enhances the precision of $\cos\delta_{CP}$, whereas the Neutrino Factory provides superior sensitivity regarding the octant of \thetamu. The inclusion of both ESSnuSB and Neutrino Factory is complementary for the precise measurement of both $\cos\delta_{CP}$ and $\sin^2\theta_{23}$, thereby enhancing our sensitivity to test the class of flavor models discussed in Section~\ref{sec:flavormodel}.

\end{document}